\documentclass[aip,jcp,reprint,twocolumn,superscriptaddress,showpacs,floatfix]{revtex4-1}

\usepackage{epsfig}
\usepackage{mathrsfs}
\usepackage{graphicx}
\usepackage{amssymb}
\usepackage{dcolumn}
\usepackage{bm}
\usepackage{gensymb}
\usepackage{physics}
\usepackage{setspace}
\usepackage{float}
\usepackage{longtable}
\usepackage{multirow}
\usepackage{color}
\usepackage{tabularx}
\definecolor{red}{rgb}{1,0,0}
\newcommand{\beq}{\begin{equation}}
\newcommand{\eeq}{\end{equation}}

\newcommand{\ftwo}[0]{$\mathrm{F_2}$}
\newcommand{\ftwoplus}[0]{$\mathrm{F_2^+}$}
\newcommand{\ccpq}[0]{CC($P$;$Q$)}
\newcommand{\ccp}[0]{CC($P$)}

\newcommand{\ccpqT}[0]{CC($P$;$Q$)$_\mathrm{(T)}$}

\newcommand{\HP}[1]{\mathscr{H}^{(P)}(#1)}
\newcommand{\HQ}[1]{\mathscr{H}^{(Q)}(#1)}

\begin{document}

\title{Converging High-Level Coupled-Cluster Energetics via Adaptive Selection of Excitation Manifolds
Driven by Moment Expansions}

\author{Karthik Gururangan}
\affiliation{Department of Chemistry, Michigan State University, 
East Lansing, Michigan 48824, USA}

\author{Piotr Piecuch}
\thanks{Corresponding author}
\email[e-mail: ]{piecuch@chemistry.msu.edu.}
\affiliation{Department of Chemistry, Michigan State University, 
East Lansing, Michigan 48824, USA}
\affiliation{Department of Physics and Astronomy, Michigan State University, 
East Lansing, Michigan 48824, USA}


\begin{abstract}
A novel approach to rapidly converging high-level coupled-cluster (CC) energetics in an automated fashion
is proposed. The key idea is an adaptive selection of the excitation manifolds defining higher--than--two-body
components of the cluster operator inspired by the CC($P$;$Q$) moment expansions. The usefulness of the resulting
methodology is illustrated by molecular examples where the goal is to recover the electronic energies obtained
using the CC method with a full treatment of singly, doubly, and triply excited clusters (CCSDT) when the
noniterative triples corrections to CCSD fail.
\end{abstract}

\maketitle

\section{Introduction}
\label{sec1}

It is well established that size extensive approaches that rely on the exponential wave function
ansatz \cite{Hubbard:1957,Hugenholtz:1957} of coupled-cluster (CC) theory,
\cite{Coester:1958,Coester:1960,cizek1,cizek2,cizek3,cizek4}
\beq
|\Psi_{0}\rangle = e^T|\Phi\rangle,
\label{eq-ccansatz}
\eeq
where $\ket{\Phi}$ is the reference determinant defining the Fermi
vacuum and
\beq
T = \sum_{n=1}^{N}T_n
\label{eq-clusterop}
\eeq
is the cluster operator, with $T_{n}$ designating its $n$-body
($n$-particle--$n$-hole or $n{\rm p}\mbox{-}n{\rm h}$) component and $N$ the number of correlated
fermions, and their extensions to excited, open-shell, and multiconfigurational states
are among the best ways of tackling many-body correlations in
atoms, molecules, and nuclei.\cite{paldus-li,crawford-schaefer,bartlett-musial2007,kum78,nucbook1,hagen2014}
This is especially true in studies of structural and spectroscopic properties of molecular systems,
chemical reaction pathways, noncovalent interactions, and photochemistry. In all
of these and similar cases, the CC hierarchy, including CCSD, where $T$ is truncated at
$T_{2}$,\cite{ccsd,ccsd2} CCSDT, where $T$ is truncated at $T_{3}$,\cite{ccfullt,ccfullt2}
CCSDTQ, where $T$ is truncated at $T_{4}$,\cite{ccsdtq0,ccsdtq2} etc., and its equation-of-motion
(EOM) \cite{emrich,eomcc3,eomccsdt1,eomccsdt3,kallaygauss,hirata1} and linear-response
\cite{monk,monk2,mukherjee_lrcc,sekino-rjb-1984,lrcc3,lrcc4,jorgensen,kondo-1995,kondo-1996} extensions
rapidly converge to the exact, full configuration interaction (FCI) limit \cite{bartlett-musial2007}
[as usual, letters S, D, T, and Q stand for single (1p-1h), double (2p-2h), triple (3p-3h), and quadruple (4p-4h)
excitations relative to $\ket{\Phi}$, respectively].
The problem is that the $\mathscr{N}^{8}$ computational steps of CCSDT or the $\mathscr{N}^{10}$ operational
count of CCSDTQ, where $\mathscr{N}$ is a measure of the system size, make such high-level CC
approaches, needed to achieve a quantitative description,
prohibitively expensive in the majority of applications of interest. Thus, one of the
biggest challenges in the CC theory has been to design numerically efficient
ways of incorporating higher--than--two-body components of the cluster operator $T$ and, in the case
of excited states, excitation operator
\beq
R_{\mu} = r_{\mu,0} {\bf 1} + \sum_{n=1}^{N} R_{\mu,n}
\label{eq-excitationop}
\eeq
entering the EOMCC wave function ansatz
\beq
|\Psi_{\mu}\rangle = R_{\mu} e^T|\Phi\rangle,
\label{eq-eomccansatz}
\eeq
where $R_{\mu,n}$
is the $n{\rm p}\mbox{-}n{\rm h}$ component of $R_{\mu}$ and ${\bf 1}$ is the unit operator, capable of
reducing the enormous costs of CCSDT, CCSDTQ, EOMCCSDT,\cite{eomccsdt1,eomccsdt3} and similar computations,
while eliminating failures of the popular perturbative approximations, such as CCSD(T),\cite{ccsdpt}
CCSDT-1,\cite{ref:21a,ref:21b} CC3,\cite{cc3_2} or CCSD(TQ$_{\rm f}$),\cite{ccsdtq-f} which
are more practical, but produce large and often uncontrollable errors in chemical bond stretching
and other multireference situations.\cite{paldus-li,bartlett-musial2007,irpc,PP:TCA,piecuch-qtp}
Solid-state applications of the periodic CC framework \cite{berkelbach2017,berkelbach2021,gruneis2019,gruneis2021}
and nuclear structure CC computations \cite{nucbook1,hagen2014,nuclei1,nuclei5,nuclei12}
face similar challenges.

In this article, we propose a novel approach to converging high-level CC energetics of the CCSDT, CCSDTQ,
and similar types, even in cases of stronger correlations, such as those characterizing stretched
chemical bonds and biradical species, where higher--than--two-body components of the cluster operator
$T$ become large, in a fully automated, ``black-box'', manner and at small fractions of the
computational costs. The key idea is an adaptive selection of the leading determinants or excitation
amplitude types needed to define the $T_{n}$ (in the case of excited states, also $R_{\mu,n}$)
components with $n > 2$, which takes advantage of the intrinsic structure of the
moment expansions abbreviated as CC($P$;$Q$), and the associated {\it a posteriori} energy
corrections to capture the remaining correlations, without any reference to the user- and
system-dependent or non-CC concepts adopted in our previous CC($P$;$Q$) work.
\cite{jspp-chemphys2012,jspp-jcp2012,jspp-jctc2012,nbjspp-molphys2017,ccpq-be2-jpca-2018,ccpq-mg2-mp-2019,%
stochastic-ccpq-prl-2017,stochastic-ccpq-molphys-2020,stochastic-ccpq-jcp-2021,cipsi-ccpq-2021,%
stochastic-ccpq-biradicals-jcp-2022} Using molecular examples characterized by substantial electronic
quasidegeneracies and large and highly nonperturbative $T_{3}$ components, we show that the proposed
adaptive, self-improving, CC($P$;$Q$) methodology using excitation spaces truncated at triples
recovers the parent CCSDT energetics with a relatively small computational effort which is
not very far removed from the noniterative triples corrections to CCSD defining
CCSD(T) and its completely renormalized CR-CC(2,3) extension,\cite{crccl_jcp,crccl_cpl,crccl_molphys,%
crccl_jpc,crccl_ijqc} even when these corrections fail or struggle (for other attempts to develop
adaptive CC ideas, see Ref.\ \onlinecite{adaptiveCC-2010}).

\section{Theory and Algorithmic Details}
\label{sec2}

\subsection{Key elements of the CC(\protect\mbox{\boldmath$P$;$Q$}) formalism}
\label{sec2.1}

We begin with a summary of the CC($P$;$Q$) formalism, as applied to many-electron systems, which was
originally formulated in Refs.\ \onlinecite{jspp-chemphys2012,jspp-jcp2012}. Each
CC($P$;$Q$) calculation begins by identifying two disjoint subspaces of the $N$-electron Hilbert space,
referred to as the $P$ and $Q$ spaces. The former space, designated as $\mathscr{H}^{(P)}$, is
spanned by the excited determinants $|\Phi_K\rangle = E_K|\Phi\rangle$, where $E_K$ is the elementary
particle--hole excitation operator generating $|\Phi_K\rangle$ from $|\Phi\rangle$, which, together with
$|\Phi\rangle$, dominate the ground state $|\Psi_{0}\rangle$ or the ground ($\mu = 0$) and excited ($\mu > 0$) states
$|\Psi_{\mu}\rangle$ of the $N$-electron system of interest. The latter space, designated as $\mathscr{H}^{(Q)}$
[${\mathscr H}^{(Q)} \subseteq ({\mathscr H}^{(0)} \oplus {\mathscr H}^{(P)})^{\perp}$,
where ${\mathscr H}^{(0)}$ is a one-dimensional subspace spanned by the reference determinant $|\Phi\rangle$],
is used to construct the noniterative corrections $\delta_{\mu}(P;Q)$ to the energies $E_{\mu}^{(P)}$ obtained
in the CC/EOMCC calculations in the $P$ space, abbreviated as CC($P$) for the ground state and EOMCC($P$) for
excited states, due to the higher-order correlation effects not captured by CC($P$)/EOMCC($P$).

Once the $P$ and $Q$ spaces are defined, we proceed as follows. First, we solve the CC($P$) equations,
\beq
\mathfrak{M}_{0,K}(P) = 0, \;\; |\Phi_K\rangle \in \mathscr{H}^{(P)},
\label{eq:cceqs}
\eeq
to determine the truncated form of the
cluster operator $T$ corresponding to the content of the $P$ space,
\beq
T^{(P)} = \sum_{\ket{\Phi_K} \in \mathscr{H}^{(P)}} t_K E_K,
\label{eq:tp}
\eeq
where $t_{K}$s are the relevant cluster
amplitudes, and the CC($P$) ground-state energy
\beq
E_{0}^{(P)} = \langle \Phi | \overline{H}^{(P)} | \Phi \rangle.
\label{eq:ccpenergy}
\eeq
Here,
\beq
\mathfrak{M}_{0,K}(P) = \langle \Phi_K | \overline{H}^{(P)}|\Phi\rangle,
\label{eq:mom}
\eeq
with
\beq
\overline{H}^{(P)} = e^{-T^{(P)}}He^{T^{(P)}},
\label{eq:hbar}
\eeq
are the generalized moments of the CC($P$) equations.
\cite{moments,leszcz,ren1} We also determine the hole--particle deexcitation operator
\beq
L_{0}^{(P)} = {\bf 1} +  \sum_{\ket{\Phi_K} \in \mathscr{H}^{(P)}} l_{0,K} (E_K)^{\dagger},
\label{eq:lzerop}
\eeq
which defines the bra ground state $\langle \tilde{\Psi}_{0}^{(P)}| = \langle \Phi | L_{0}^{(P)} e^{-T^{(P)}}$
associated with the CC($P$) ket state $|\Psi_{0}^{(P)} \rangle = e^{T^{(P)}}|\Phi\rangle$ satisfying
the normalization condition $\langle \tilde{\Psi}_{0}^{(P)} | \Psi_{0}^{(P)}\rangle = 1$,
by solving the linear system
\beq
\bra{\Phi} L_{0}^{(P)} \overline{H}^{(P)} \ket{\Phi_K} = E_{0}^{(P)} l_{0,K}, \;\;
\ket{\Phi_K}\in\mathscr{H}^{(P)},
\label{left-cc-gen}
\eeq
where $E_{0}^{(P)}$ is the CC($P$) ground-state energy obtained with Eq. (\ref{eq:ccpenergy}).
If there is interest in the ground- as well as excited-state energetics, we follow the CC($P$) calculations
by the diagonalization of the similarity-transformed Hamiltonian $\overline{H}^{(P)}$, Eq. (\ref{eq:hbar}),
in the $P$ space
$\mathscr{H}^{(P)}$ to obtain the EOMCC($P$) energies $E_{\mu}^{(P)}$ and the corresponding
EOM excitation and deexcitation operators,
\beq
R_{\mu}^{(P)} = r_{\mu,0} {\bf 1} + \sum_{\ket{\Phi_K} \in \mathscr{H}^{(P)}} r_{\mu,K} E_K
\label{eq:rmup}
\eeq
and
\beq
L_{\mu}^{(P)} = \delta_{\mu,0} {\bf 1} + \sum_{\ket{\Phi_K} \in \mathscr{H}^{(P)}} l_{\mu,K} (E_K)^{\dagger},
\label{eq:lmup}
\eeq
respectively, where the amplitudes $r_{\mu,K}$, along with $r_{\mu,0}$, define the EOMCC($P$) ket states
$|\Psi_{\mu}^{(P)}\rangle = R_{\mu}^{(P)} e^{T^{(P)}}|\Phi\rangle$ and their left-eigenstate $l_{\mu,K}$ 
counterparts define the EOMCC($P$) bra states $\langle \tilde{\Psi}_{\mu}^{(P)} | = \langle \Phi|
L_{\mu}^{(P)} e^{-T^{(P)}}$ satisfying the biorthonormality condition
$\langle\tilde{\Psi}_{\mu}^{(P)}|\Psi_{\nu}^{(P)}\rangle = \langle\Phi|L_{\mu}^{(P)} R_{\nu}^{(P)}|\Phi\rangle
= \delta_{\mu,\nu}$.
Once $T^{(P)}$, $L_{0}^{(P)}$, and $E_{0}^{(P)}$ and, in the case of excited states, $R_{\mu}^{(P)}$,
$L_{\mu}^{(P)}$, and $E_{\mu}^{(P)}$ are known, we proceed to the final step, which
is the determination of the CC($P$;$Q$) energies
\begin{equation}
E_{\mu}^{(P+Q)} = E_{\mu}^{(P)} + \delta_{\mu}(P;Q) ,
\label{eq:ccpq_energy}
\end{equation}
where corrections $\delta_{\mu}(P;Q)$ are given by
\begin{equation}
\delta_{\mu}(P;Q) = \sum_{\ket{\Phi_K} \in \mathscr{H}^{(Q)}} \ell_{\mu,K}(P) \: \mathfrak{M}_{\mu,K}(P) .
\label{eq:mmcorrection}
\end{equation}
The ground-state moments $\mathfrak{M}_{0,K}(P)$, Eq. (\ref{eq:mom}), and their excited-state counterparts
\cite{kkppeom2,kkppeom3,kkppeom}
\beq
\mathfrak{M}_{\mu,K}(P) = \langle \Phi_K | \overline{H}^{(P)} R_{\mu}^{(P)} |\Phi\rangle
\label{eq:momex}
\eeq
entering Eq. (\ref{eq:mmcorrection}) correspond to projections of the CC($P$) and EOMCC($P$)
equations on the $Q$-space determinants $\ket{\Phi_K} \in \mathscr{H}^{(Q)}$. The coefficients
$\ell_{\mu,K}(P)$ that multiply these moments are obtained using the Epstein--Nesbet-like formula
\beq
\ell_{\mu,K}(P) = \langle \Phi| L_{\mu}^{(P)} \overline{H}^{(P)} |\Phi_K \rangle / D_{\mu,K}^{(P)},
\label{eq:ell}
\eeq
where
\beq
D_{\mu,K}^{(P)} = E_{\mu}^{(P)} - \langle \Phi_K | \overline{H}^{(P)} | \Phi_K \rangle.
\label{eq:denom}
\eeq
One can replace the denominators $D_{\mu,K}^{(P)}$, Eq. (\ref{eq:denom}),
by their M{\o}ller--Plesset-like analogs, but,
as shown in
Refs.\ \onlinecite{jspp-jctc2012,nbjspp-molphys2017,stochastic-ccpq-prl-2017,stochastic-ccpq-jcp-2021,%
crccl_jcp,crccl_cpl,crccl_jpc,crccl_ijqc2}, the Epstein--Nesbet-like form of $D_{\mu,K}^{(P)}$
results in a more accurate description.

The CC($P$;$Q$) formalism can be viewed as
a generalization of the biorthogonal moment expansions of Refs.\ \onlinecite{crccl_jcp,crccl_cpl,%
crccl_molphys,crccl_ijqc,crccl_ijqc2} and, as such, includes the aforementioned CR-CC(2,3) approach
and its higher-order \cite{ptcp2007,msg65,nuclei8,nbjspp-molphys2017,ccpq-be2-jpca-2018}
and excited-state \cite{crccl_molphys,crccl_ijqc2,7hq} extensions, but its key advantage,
which the previous moment expansions did not have, is the possibility of
making unconventional choices of the $P$ and $Q$ spaces and relaxing
the lower-order $T_{n}$ and $R_{\mu,n}$ components with $n \leq 2$ in the presence
of their higher--than--two-body counterparts, such as $T_{3}$ and $R_{\mu,3}$, which the
CCSD(T), CR-CC(2,3), and other triples or higher-order corrections to CCSD or EOMCCSD cannot do.
There are important problems, including the molecular examples considered in this work and many
others, where the decoupling of the higher--than--two-body components of $T$ and $R_{\mu}$
from their lower-order $T_{n}$ and $R_{\mu,n}$ counterparts with $n \leq 2$,
characterizing CCSD(T), CR-CC(2,3), and similar methods, results in large errors
in the calculated potential surfaces and activation barriers,
\cite{jspp-chemphys2012,jspp-jcp2012,nbjspp-molphys2017,stochastic-ccpq-prl-2017,stochastic-ccpq-jcp-2021,%
cipsi-ccpq-2021} singlet--triplet gaps,\cite{jspp-jctc2012,stochastic-ccpq-biradicals-jcp-2022}
and electronic excitation spectra.\cite{stochastic-ccpq-molphys-2020} As shown in Refs.\
\onlinecite{jspp-chemphys2012,jspp-jcp2012,jspp-jctc2012,nbjspp-molphys2017,ccpq-be2-jpca-2018,ccpq-mg2-mp-2019,%
stochastic-ccpq-prl-2017,stochastic-ccpq-molphys-2020,stochastic-ccpq-jcp-2021,cipsi-ccpq-2021,%
stochastic-ccpq-biradicals-jcp-2022}, the CC($P$;$Q$) formalism addresses this issue and is capable of
recovering the high-level CC/EOMCC energetics of the CCSDT,
\cite{jspp-chemphys2012,jspp-jcp2012,jspp-jctc2012,nbjspp-molphys2017,ccpq-be2-jpca-2018,ccpq-mg2-mp-2019,%
stochastic-ccpq-prl-2017,stochastic-ccpq-molphys-2020,stochastic-ccpq-jcp-2021,cipsi-ccpq-2021,%
stochastic-ccpq-biradicals-jcp-2022}
CCSDTQ,\cite{nbjspp-molphys2017,ccpq-be2-jpca-2018,ccpq-mg2-mp-2019,stochastic-ccpq-jcp-2021}
and EOMCCSDT \cite{stochastic-ccpq-molphys-2020} types at small
fractions of the computational costs, even when electronic quasidegeneracies become substantial,
by incorporating
the dominant
triply or triply and quadruply excited determinants in the $P$ space, while using corrections
$\delta_{\mu}(P;Q)$ to account for the remaining correlations of the parent CC/EOMCC
approach of interest, but in order to do this one has to come up with a method of identifying the leading
higher--than--doubly excited determinants for inclusion in the $P$ space. So far, this has been
accomplished with the help of user-specified active orbitals,
resulting in the CC(t;3), CC(t,q;3), CC(t,q;3,4), etc. hierarchy,
\cite{jspp-chemphys2012,jspp-jcp2012,jspp-jctc2012,nbjspp-molphys2017,ccpq-be2-jpca-2018,ccpq-mg2-mp-2019}
Quantum Monte Carlo (QMC) wave function propagations in the many-electron Hilbert space employing the CIQMC
\cite{Booth2009,Cleland2010,fciqmc-uga-2019,ghanem_alavi_fciqmc_jcp_2019,ghanem_alavi_fciqmc_2020}
and CCMC \cite{Thom2010,Franklin2016,Spencer2016,Scott2017} approaches, resulting in the semi-stochastic
CC($P$;$Q$) theories \cite{stochastic-ccpq-prl-2017,stochastic-ccpq-molphys-2020,stochastic-ccpq-jcp-2021,%
stochastic-ccpq-biradicals-jcp-2022} (cf., also, Ref.\ \onlinecite{eomccp-jcp-2019}), and sequences of
Hamiltonian diagonalizations in systematically grown subspaces of the many-electron Hilbert space
originating from one of the selected CI schemes,
\cite{sci_1,sci_2,sci_3,sci_4,adaptive_ci_1,adaptive_ci_2,asci_1,asci_2,ici_1,ici_2,shci_1,shci_2,shci_3,
cipsi_1,cipsi_2} abbreviated as CIPSI,\cite{sci_3,cipsi_1,cipsi_2}
resulting in the CIPSI-driven CC($P$;$Q$) algorithm.\cite{cipsi-ccpq-2021} All of these attempts to
design the appropriate $P$ and $Q$ spaces for accurate CC($P$;$Q$) computations
have had their successes, but the adaptive, self-improving,
CC($P$;$Q$) framework, which we discuss next, takes us to an entirely new level by freeing us from the
user-defined active orbitals and non-CC (CIQMC, CIPSI) or stochastic (CIQMC, CCMC) concepts
adopted in our previous CC($P$;$Q$) work.

\subsection{Adaptive CC(\protect\mbox{\boldmath$P$;$Q$}): General considerations}
\label{sec2.2}

In developing the adaptive CC($P$;$Q$) methodology, we have been inspired by the modern
CIPSI algorithm described in Refs.\ \onlinecite{cipsi_1,cipsi_2}, available in the open-source
Quantum Package 2.0,\cite{cipsi_2} which we exploited in Refs.\ \onlinecite{cipsi-ccpq-2021,eccc-jctc-2021}.
As already alluded to above, the main idea of CIPSI is a series of Hamiltonian diagonalizations in increasingly
large, iteratively defined, subspaces of the many-electron Hilbert space, which are followed by correcting the
resulting energies using expressions originating from the second-order many-body perturbation theory to
estimate the remaining correlations.\cite{sci_3,cipsi_1,cipsi_2} If $\mathcal{V}_{\mathrm{int}}^{(k)}$, where
$k = 0, 1, 2, \ldots$ enumerates the consecutive CIPSI iterations, designates the current diagonalization subspace,
the $\mathcal{V}_{\mathrm{int}}^{(k+1)}$ space for the subsequent Hamiltonian diagonalization is constructed by
arranging the candidate determinants $|\Phi_\alpha\rangle$ from outside $\mathcal{V}_{\mathrm{int}}^{(k)}$
for a potential inclusion in $\mathcal{V}_{\mathrm{int}}^{(k+1)}$ in descending order according to the absolute
values of the perturbative corrections $e_{\alpha,k}^{(2)}$ associated with them, starting from the
$|\Phi_\alpha\rangle$s characterized by the largest $|e_{\alpha,k}^{(2)}|$ contributions, moving toward those
that have smaller $|e_{\alpha,k}^{(2)}|$ values, and continuing until the number of determinants in
$\mathcal{V}_{\mathrm{int}}^{(k+1)}$ reaches a desired dimension (in the CIPSI algorithm of
Refs.\ \onlinecite{cipsi_1,cipsi_2}, until the dimension of $\mathcal{V}_{\mathrm{int}}^{(k+1)}$
exceeds that of $\mathcal{V}_{\mathrm{int}}^{(k)}$ by the user-defined factor $f > 1$). We can adopt
a similar strategy in designing $P$ spaces for the \ccpq{} computations.

Indeed, we can interpret Eq.\
(\ref{eq:mmcorrection}) as a sum of contributions $\delta_{\mu,K}(P;Q)$ due to the individual determinants
from the $Q$ space, $|\Phi_K\rangle \in \mathscr{H}^{(Q)}$, evaluated as
\begin{equation}
\delta_{\mu,K}(P;Q) = \ell_{\mu,K}(P)\: 
\mathfrak{M}_{\mu,K}(P) ,
\label{eq:deltapq_K}
\end{equation}
which, in analogy to the perturbative $e_{\alpha,k}^{(2)}$ corrections that measure the significance of the
candidate determinants $|\Phi_\alpha\rangle$ in CIPSI, determine the importance of the $Q$-space determinants
$|\Phi_K\rangle$. One can, therefore, propose an adaptive, self-improving, \ccpq{} scheme, in which we construct an
approximation to the high-level CC/EOMCC approach of interest (in the numerical examples in Sec. \ref{sec3},
CCSDT) by a series of \ccpq{} calculations using increasingly large, iteratively defined, $P$ spaces $\HP{k}$,
where $k = 0, 1, 2, \ldots$ enumerates the consecutive \ccpq{} computations, with the corresponding $Q$
subspaces $\HQ{k}$ being defined as complementary excitation spaces, such that $\HP{k} \oplus \HQ{k}$ is always
equivalent to the entire excitation manifold appropriate for the CC/EOMCC method we are targeting, independent
of $k$ (when targeting CCSDT, all singly, doubly, and triply excited determinants,
when targeting CCSDTQ, all singly, doubly, triply, and quadruply excited determinants, etc.).
The initial $P$ space $\HP{0}$
can be a conveniently chosen zeroth-order excitation manifold, such as the space of singly and doubly
excited determinants, $|\Phi_{i}^{a}\rangle$ and $|\Phi_{ij}^{ab}\rangle$, respectively, where
$i, j, \ldots$ ($a, b, \ldots$) designate the spin-orbitals occupied (unoccupied) in $|\Phi\rangle$,
and the remaining subspaces are constructed via a recursive process analogous to that used in CIPSI,
where the $P$ space  $\HP{k+1}$ is obtained by augmenting its $\HP{k}$ predecessor with a subset of the
leading $Q$-space determinants $|\Phi_K\rangle \in \HQ{k}$ [when targeting CCSDT, the leading triply
excited determinants $|\Phi_{ijk}^{abc}\rangle$ outside $\HP{k}$,
when targeting CCSDTQ, the leading triply and quadruply excited determinants outside $\HP{k}$, etc.]
identified with the help of corrections
$\delta_{\mu,K}(P;Q)$, Eq. (\ref{eq:deltapq_K}). Inspired by the CIPSI algorithm,
one can enlarge the current subspace $\HP{k}$ to construct the $\HP{k+1}$ space for the subsequent \ccpq{}
computation by arranging the candidate determinants $|\Phi_K\rangle \in \HQ{k}$ in descending order according
to the $\delta_{\mu,K}(P;Q)$ corrections associated with them, starting from the $|\Phi_K\rangle$s
characterized by the largest $|\delta_{\mu,K}(P;Q)|$ contributions, moving toward those that have
smaller $|\delta_{\mu,K}(P;Q)|$ values, and continuing until the total number of the $Q$-space determinants
in $\HP{k+1}$ reaches a certain fraction of all higher-rank determinants relevant to the
CC/EOMCC approach of interest (e.g., triples when targeting CCSDT
or triples and quadruples when targeting CCSDTQ).
Clearly, the adaptive \ccpq{} procedure,
as described above, guarantees convergence toward the high-level CC/EOMCC approach it targets, but,
following the computational cost analysis of the \ccpq{}
methods
\cite{stochastic-ccpq-prl-2017,stochastic-ccpq-molphys-2020,stochastic-ccpq-jcp-2021}
(cf., also, Refs.\
\onlinecite{jspp-chemphys2012,jspp-jcp2012,jspp-jctc2012,nbjspp-molphys2017,ccpq-be2-jpca-2018,ccpq-mg2-mp-2019,%
cipsi-ccpq-2021,stochastic-ccpq-biradicals-jcp-2022}), in order to be an attractive 
approach,
it has to be
capable of recovering the target CC/EOMCC energetics to a very good accuracy with small fractions
of higher-rank determinants relevant to the CC/EOMCC method of interest in the underlying $P$ spaces.

\subsection{Adaptive CC(\protect\mbox{\boldmath$P$;$Q$}) and CC(\protect\mbox{\boldmath$P$;$Q$})$_{\textbf{(T)}}$
approaches aimed at CCSDT and their computational implementation}
\label{sec2.3}

To illustrate the benefits offered by the approach proposed in Sec. \ref{sec2.2},
we wrote an adaptive \ccpq{} code aimed at recovering
the ground-state CCSDT energetics, which we incorporated in our open-source CCpy package available on GitHub,
\cite{CCpy-GitHub} implemented in Python and using Numpy and Fortran
in the computationally critical parts. In this case, we apply Eq. (\ref{eq:mmcorrection})
to a situation where the $k$th $P$ space $\HP{k}$ is
spanned by all singly and doubly excited determinants and a subset of triply excited determinants identified by
the adaptive \ccpq{} algorithm and the associated $Q$ space $\HQ{k}$ consists of the remaining triples
not included in $\HP{k}$, i.e.,
\begin{equation}
\delta_{0}(P^{(k)};Q^{(k)}) = 
\sum_{|\Phi_{ijk}^{abc}\rangle \in \HQ{k}} \delta_{ijk,abc}(k) ,
\label{eq:delta_k}
\end{equation}
where
\begin{equation}
\delta_{ijk,abc}(k) = \ell_{ijk}^{abc}(P^{(k)})\:
\mathfrak{M}_{abc}^{ijk}(P^{(k)})
\label{eq:delta_ijkabc}
\end{equation}
is the contribution to $\delta_0(P^{(k)};Q^{(k)})$ that corresponds to a given triply excited determinant
$|\Phi_{ijk}^{abc}\rangle \in \HQ{k}$. The $\ell_{ijk}^{abc}(P^{(k)})$ and $\mathfrak{M}_{abc}^{ijk}(P^{(k)})$
quantities in Eq. (\ref{eq:delta_ijkabc}) are the coefficients $\ell_{0,K}(P)$ and moments
$\mathfrak{M}_{0,K}(P)$ entering the ground-state \ccpq{} correction $\delta_{0}(P;Q)$,
Eq. (\ref{eq:mmcorrection}) in which $\mu$ is set at 0,
adapted to the above definitions of the $\HP{k}$ and $\HQ{k}$ spaces. For clarity of this description,
the $P$ and $Q$ symbols seen in Eqs. (\ref{eq:delta_k}) and (\ref{eq:delta_ijkabc}), which represent the
$\HP{k}$ and $\HQ{k}$ spaces used in the $k$th iteration of the adaptive \ccpq{} procedure, are labeled
with the additional superscript $(k)$.

In our current implementation of the adaptive \ccpq{} approach
aimed at recovering the CCSDT-level energetics, we invoked the so-called two-body approximation that
was successfully used in some of our earlier \ccpq{}-related work.
\cite{jspp-chemphys2012,jspp-jcp2012,jspp-jctc2012,nbjspp-molphys2017,eccc-jctc-2021} This means that we
replaced moments $\mathfrak{M}_{abc}^{ijk}(P^{(k)})$ and coefficients $\ell_{ijk}^{abc}(P^{(k)})$ entering
Eq. (\ref{eq:delta_ijkabc}) by $\langle \Phi_{ijk}^{abc} | \overline{H}^{(P)}(2) | \Phi \rangle$ and
$\langle \Phi | ({\bf 1} + L_{0,1} + L_{0,2}) \: \overline{H}^{(P)}(2) | \Phi_{ijk}^{abc} \rangle/
(E_{0}^{(P)} - \langle \Phi_{ijk}^{abc} | \overline{H}^{(P)}(2) | \Phi_{ijk}^{abc} \rangle)$, respectively,
where $\overline{H}^{(P)}(2) = e^{-T_{1} - T_{2}} H e^{T_{1} + T_{2}}$, with $T_{1}$ and $T_{2}$ designating
the one- and two-body components of the cluster operator $T^{(P)}$ obtained in the CC($P$) calculations in
$\HP{k}$, is an approximation to the true similarity-transformed Hamiltonian $\overline{H}^{(P)} =
e^{-T_{1} - T_{2} - T_{3}^{(P)}} H e^{T_{1} + T_{2} + T_{3}^{(P)}}$, where $T_{3}^{(P)}$ is a three-body
component of $T^{(P)}$. The one- and two-body deexcitation operators $L_{0,1}$ and  $L_{0,2}$, which enter the
formula for the $\ell_{ijk}^{abc}(P^{(k)})$ coefficients used in our adaptive \ccpq{} code, are obtained by
solving the left eigenvalue problem involving $\overline{H}^{(P)}(2)$ in the space spanned by singly and doubly
excited determinants. Thus, in analogy to $T^{(P)}$, the three-body component $L_{0,3}^{(P)}$ of the deexcitation
operator $L_{0}^{(P)}$ corresponding to the $P$ space $\HP{k}$ is neglected. In this way, we preserve the philosophy
of the \ccpq{} algorithm and account for the relaxation of $T_{1}$, $T_{2}$, $L_{0,1}$, and $L_{0,2}$
in the presence of the three-body components of $T^{(P)}$ obtained in the CC($P$) calculations in $\HP{k}$,
while avoiding the more complex computational steps associated with the use of the complete form of
$\overline{H}^{(P)}$ in determining the $\delta_{ijk,abc}(k)$ corrections.

In the adaptive \ccpq{} code that we incorporated in the CCpy package available on GitHub,\cite{CCpy-GitHub} the
$k=0$ $P$ space $\HP{0}$, used to initiate the calculations, is defined as all singly and doubly excited
determinants and the corresponding $Q$ space $\HQ{0}$ consists of all triples. This means that the $k=0$ CC($P$)
and \ccpq{} energies are identical to those obtained with CCSD and CR-CC(2,3), respectively. We then follow the
recursive procedure described in Sec. \ref{sec2.2} by moving more and more triply excited determinants from the $Q$
to $P$ spaces. We enlarge the $k$th subspace $\HP{k}$, which consists of all singles and doubles and,
when $k > 0$, the subset of triples identified in the previous iteration, to construct the $\HP{k+1}$ space
for the subsequent \ccpq{} computation by arranging the candidate triply excited determinants
$|\Phi_{ijk}^{abc}\rangle \in \HQ{k}$ in descending order according to the
$\delta_{ijk,abc}(k)$
corrections associated with them, starting from the triples characterized by the largest
$|\delta_{ijk,abc}(k)|$
contributions, moving toward those that have smaller
$|\delta_{ijk,abc}(k)|$
values, and continuing until the total number of triply
excited determinants in $\HP{k+1}$ represents an increase of the number of triples in $\HP{k}$ by 1\%.
We could have chosen any other incremental fraction of triples when enlarging the $\HP{k}$,
$k = 0, 1, 2, \ldots \,$, spaces, but, given our desire to use as small fractions of triples in the consecutive
CC($P$) runs as possible, so that the iterative CC($P$) steps preceding the determination of the
$\delta_{0}(P^{(k)};Q^{(k)})$ corrections are much less expensive than those of the CCSDT target,
we felt that 1\% is a good incremental fraction to start with in our preliminary numerical tests
reported in this initial study (we will explore
other incremental fractions of triples in the future work). The algorithm that takes full
advantage of the speedups offered by the adaptive \ccpq{} procedure relative
to its CCSDT parent, when the underlying $P$ space contains a small fraction of triply excited determinants,
along with other information about the CCpy package in which our adaptive \ccpq{} approach targeting CCSDT
has been implemented, will be discussed in a separate publication.

In the adaptive \ccpq{} computations reported
in this work, we distinguish between the relaxed and unrelaxed schemes. In the relaxed variant of the
adaptive \ccpq{}, we solve the \ccp{} equations for the singly, doubly, and triply excited cluster amplitudes
corresponding to the content of each $\HP{k}$ space and recompute the corresponding triples corrections
$\delta_{ijk,abc}(k)$
accordingly, increasing the number of triples, when going from $\HP{k}$
to $\HP{k+1}$, by 1\%. In the unrelaxed \ccpq{} approach, we simply
pick a particular fraction of triples for inclusion in the $P$ space (say, 3\% or 5\%) that have the
largest initial
$|\delta_{ijk,abc}(0)|$
values determined using the $T_1$ and $T_2$ amplitudes
obtained with CCSD [i.e., the largest absolute values of the
$\delta_{ijk,abc}(0)$
contributions to the triples correction of CR-CC(2,3)] and solve the \ccp{} equations for the singly,
doubly, and triply excited cluster amplitudes with this particular fraction of triples, adding the
correction due to the missing $T_3$ correlations using Eq. (\ref{eq:delta_k}). The relaxed
variant of the adaptive \ccpq{} approach has an advantage of approaching the target CCSDT energetics
as $k \rightarrow \infty$, but the unrelaxed scheme, which does not require an iterative construction of
multiple $\HP{k}$ spaces,
is less expensive. One of the goals of this initial study of the adaptive
\ccpq{} approach is to compare the two schemes.

Given the straightforward relationship between the triples corrections of CR-CC(2,3) and CCSD(T), discussed in
Refs.\ \onlinecite{crccl_jcp,crccl_cpl,ptcp2007,crccl_jpc,jspp-chemphys2012}, and the fact that the
$k=0$ \ccpq{} energies are identical to those obtained with CR-CC(2,3), we also implemented the CCSD(T)-like
variants of the relaxed and unrelaxed adaptive \ccpq{} algorithms, abbreviated as \ccpqT{}. One obtains the
adaptive \ccpqT{} approach by replacing moments $\mathfrak{M}_{abc}^{ijk}(P^{(k)})$ and coefficients
$\ell_{ijk}^{abc}(P^{(k)})$ in Eq. (\ref{eq:delta_ijkabc}) by $\langle \Phi_{ijk}^{abc} | V_{N} T_{2} |\Phi \rangle$
and $\langle \Phi | (T_{1}^{\dagger} + T_{2}^{\dagger}) V_{N} | \Phi_{ijk}^{abc} \rangle/
(\epsilon_{i} + \epsilon_{j} + \epsilon_{k} - \epsilon_{a} - \epsilon_{b} - \epsilon_{c})$, respectively,
where $V_{N}$ is the two-body part of the normal-ordered Hamiltonian, $T_{1}$ and $T_{2}$
are the singly and doubly excited components of the cluster operator $T^{(P)}$ resulting from the
CC($P$) calculations in $\HP{k}$, and $\epsilon_{p}$ is the single-particle energy associated with
the spin-orbital $|p\rangle$ (the diagonal element of the Fock matrix associated with $|p\rangle$).
Since the initial $P$ space $\HP{0}$ is spanned by all singly and doubly excited determinants
and the associated $Q$ space $\HQ{0}$ contains all triples, the $k=0$ \ccpqT{} energies are identical
to those of CCSD(T). One of the objectives of this study is to determine if we could replace
the adaptive \ccpq{} approach by its CCSD(T)-like \ccpqT{} analog, which
uses the somewhat less expensive triples correction $\delta_{0}(P^{(k)};Q^{(k)})$ and does not require
solving the left eigenvalue problem involving $\overline{H}^{(P)}(2)$, without
losing the accuracy of its CR-CC(2,3)-like parent.

\section{Numerical Examples}
\label{sec3}

\subsection{Molecular systems and the reasons for their choice}
\label{sec3.1}

To test the performance of the relaxed and unrelaxed variants of the adaptive \ccpq{} and \ccpqT{} approaches
based on Eqs. (\ref{eq:delta_k}) and (\ref{eq:delta_ijkabc}), especially their ability to recover or accurately
approximate the parent CCSDT energetics when the noniterative triples corrections to CCSD, represented in this
work by CCSD(T) and CR-CC(2,3), struggle, we applied them to the significantly stretched \ftwo{} and \ftwoplus{}
molecules and the reactant (R) and transition-state (TS) species involved in the automerization of cyclobutadiene,
along with the corresponding barrier height, for which full CCSDT, as shown, for example, in Refs.\
\onlinecite{jspp-chemphys2012,jspp-jcp2012,stochastic-ccpq-prl-2017,stochastic-ccpq-jcp-2021,cipsi-ccpq-2021,%
crccl_jcp,crccl_cpl,f2bh,balkova1994,ruedenberg-f2}, works well. In the case of the fluorine molecule and its cation
(Tables \ref{tab:table1} and \ref{tab:table2}), for which we used the cc-pVTZ basis set,\cite{ccpvnz} the
respective F--F bond lengths $r$ were stretched to $2r_e$, where $r_{e}$ represents the equilibrium geometry
(2.66816 bohr for \ftwo{} and 2.49822 bohr for \ftwoplus{}). The geometries of the $D_{2h}$-symmetric R and
$D_{4h}$-symmetric TS structures adopted in our
calculations for cyclobutadiene (Tables \ref{tab:table3} and \ref{tab:table4}), in which we employed the
cc-pVDZ basis,\cite{ccpvnz} were taken from Ref.\ \onlinecite{MR-AQCC}, where they were optimized using the
multireference average-quadratic CC (MR-AQCC) approach.\cite{mraqcc1,mraqcc2} With the exception of CCSDT,
all CC calculations reported in this work, including CCSD, CCSD(T), and CR-CC(2,3) and the adaptive
\ccpq{} and \ccpqT{} runs, were performed using the aforementioned
CCpy package.\cite{CCpy-GitHub}
In the case of CCSDT, we used our highly efficient in-house code, written in Fortran,
developed in the context of our previous active-orbital-based \ccpq{} studies in Refs.\
\onlinecite{jspp-chemphys2012,jspp-jcp2012,nbjspp-molphys2017}. The computations preceding the CC steps,
including the generation of one- and two-electron integrals, the restricted Hartree--Fock (RHF; \ftwo{}
and cyclobutadiene) and restricted open-shell Hartree--Fock (ROHF; \ftwoplus{}) self-consistent-field (SCF)
calculations, and the relevant integral transformations were carried out with the GAMESS package.
\cite{gamess,gamess2020}
In all post-SCF
calculations, the core orbitals correlating with the 1s shells of the F and C atoms were
kept frozen.

We chose the significantly stretched fluorine molecule and its cation with $r = 2r_{e}$, since they
are characterized by the enormous and highly nonperturbative
$T_3$ effects, which are estimated, by forming the differences of the CCSDT and CCSD energies, at about
$-63$ and $-76$ millihartree, respectively, when the cc-pVTZ basis set is employed (see Tables \ref{tab:table1}
and \ref{tab:table2}). As shown, for example, in Ref.\ \onlinecite{jspp-chemphys2012}, and as further discussed in
Refs.\ \onlinecite{stochastic-ccpq-jcp-2021,cipsi-ccpq-2021},
the $T_3$ effects in the stretched \ftwo{} and \ftwoplus{}
molecules at $r = 2r_{e}$ are so large that they exceed (\ftwo{}) or are not much smaller than (\ftwoplus{}) the
depths of the corresponding CCSDT potential wells (estimated in the case of the cc-pVTZ basis at about 57 and 116
millihartree, respectively\cite{jspp-chemphys2012,stochastic-ccpq-jcp-2021,cipsi-ccpq-2021}). The nonperturbative
character of $T_3$ correlations in the stretched \ftwo{} and \ftwoplus{} systems considered in this study is
best illustrated by examining the large errors relative to CCSDT obtained in the CCSD(T) calculations, which
in the cc-pVTZ basis used in this work are $-26.354$ and $-8.985$ millihartree, respectively
\cite{jspp-chemphys2012,stochastic-ccpq-jcp-2021,cipsi-ccpq-2021}
(cf. Table \ref{tab:table2}). The CR-CC(2,3) triples correction to CCSD, equivalent to the initial, $k=0$, \ccpq{}
calculation, improves the CCSD(T) [$k=0$ \ccpqT{}] result for the stretched \ftwo{} molecule in a substantial
manner, but, as shown in Table \ref{tab:table1} and Refs.\ \onlinecite{jspp-chemphys2012,stochastic-ccpq-jcp-2021,%
cipsi-ccpq-2021}, the 4.254  millihartree error relative to CCSDT remains. The same table
(see, also, Ref.\ \onlinecite{jspp-chemphys2012}) demonstrates that
the $r = 2r_{e}$ \ftwoplus{} system presents an even greater challenge, since in this case the
CR-CC(2,3) approach, while placing the energy above its CCSDT parent, does not reduce the large
unsigned error relative to CCSDT characterizing  CCSD(T). In fact, the CR-CC(2,3) calculation makes
it somewhat larger [10.971 millihartree, as opposed to 8.985 millihartree obtained with CCSD(T)].
This is indicative of a substantial increase in the coupling between the lower-rank $T_{1}$ and $T_{2}$
clusters and the higher-rank $T_{3}$ component compared to the similarly stretched \ftwo{} species,
which none of the noniterative triples corrections to CCSD can capture.

In the case of cyclobutadiene, the $T_{3}$ effects, again estimated by forming the differences between the
CCSDT and CCSD energies, which are $-26.827$ millihartree for the R and $-47.979$ millihartree for the TS structures
when the cc-pVDZ basis set is employed (see Tables \ref{tab:table3} and \ref{tab:table4}), are not only very
large, but also difficult to balance, resulting in a massive, 13.274 kcal/mol, error relative to CCSDT in the
CCSD value of the barrier height characterizing the automerization process, which exceeds the CCSDT activation
energy of 7.627 kcal/mol by more than 170\%.
\cite{jspp-jcp2012,stochastic-ccpq-prl-2017,stochastic-ccpq-jcp-2021,cipsi-ccpq-2021} Furthermore, the more
challenging TS structure is characterized by a large coupling of the lower-rank $T_{1}$ and $T_{2}$
components with their higher-rank $T_{3}$ counterpart, so large that none of the noniterative triples corrections
to CCSD provide a reasonable description of the barrier height.\cite{jspp-jcp2012,balkova1994,tailored3}
This, in particular, applies to the CR-CC(2,3) approach, which places the activation barrier at 16.280 kcal/mol
when the cc-pVDZ basis set is employed, instead of less than 8 kcal/mol obtained with CCSDT
\cite{jspp-jcp2012,stochastic-ccpq-prl-2017,stochastic-ccpq-jcp-2021,cipsi-ccpq-2021} (see
Table \ref{tab:table3}). Once again, CR-CC(2,3) does not improve the results of the CCSD(T) calculations,
which give a similarly poor barrier height (15.832 kcal/mol; see Table \ref{tab:table4}). The failure of the
CR-CC(2,3) and CCSD(T) methods to provide accurate activation energies for the automerization of cyclobutadiene
is a consequence of the poor performance of both approaches in the calculations for the TS species. As shown in
Refs.\ \onlinecite{jspp-jcp2012,stochastic-ccpq-prl-2017,stochastic-ccpq-jcp-2021,cipsi-ccpq-2021}
and Tables \ref{tab:table3}
and \ref{tab:table4}, the difference between the CR-CC(2,3) and CCSDT energies at the TS geometry obtained with
the cc-pVDZ basis, of 14.636 millihartree, and the analogous 14.198 millihartree difference between the CCSD(T)
and CCSDT energy values are much larger than the $\sim \!\! 1$ millihartree errors relative to CCSDT characterizing
the CR-CC(2,3) and CCSD(T) calculations for the R species.

\subsection{Performance of the adaptive CC(\protect\mbox{\boldmath$P$;$Q$}) and
CC(\protect\mbox{\boldmath$P$;$Q$})$_{\textbf{(T)}}$ approaches: Energetics}
\label{sec3.2}

In our previous studies,\cite{jspp-chemphys2012,jspp-jcp2012,stochastic-ccpq-prl-2017,stochastic-ccpq-jcp-2021,%
cipsi-ccpq-2021} we demonstrated that the active-orbital-based \ccpq{} scheme abbreviated as
CC(t;3) and the semi-stochastic, CIQMC- and CCMC-based, and CIPSI-driven \ccpq{} approaches are capable of
greatly improving the CCSD(T) and CR-CC(2,3) results for the significantly stretched \ftwo{} molecule and the
automerization of cyclobutadiene, generating the CCSDT-quality data at small fractions
of the computational effort associated with the conventional CCSDT calculations. In our initial \ccpq{} work
reported in Ref.\ \onlinecite{jspp-chemphys2012}, we showed that the CC(t;3) method can do the same for the
stretched \ftwoplus{}. We now show that the fully automated, ``black-box'', adaptive \ccpq{}
methodology, as implemented in our CCpy package,\cite{CCpy-GitHub} performs equally well, or can even be more
efficient.

Indeed, as shown in Tables \ref{tab:table1} and \ref{tab:table3}, the convergence of the adaptive \ccpq{}
calculations toward the parent CCSDT energetics is very fast. This includes the more challenging multireference
situations created by the stretched \ftwo{} and \ftwoplus{} molecules and the TS structure of cyclobutadiene,
where, as explained above, $T_{3}$ correlations are large, nonperturbative, and difficult to capture,
resulting in failures of CCSD(T) and, in the case of the latter two systems, of CR-CC(2,3), as well
as the weakly correlated cyclobutadiene R species, which the CCSD(T) and CR-CC(2,3) methods can handle,
although not perfectly. The relaxed variant of the adaptive \ccpq{} algorithm
is generally most accurate. For the most demanding cases of the stretched \ftwoplus{} and TS
species of cyclobutadiene, where the coupling of the lower-rank $T_{1}$ and $T_{2}$ clusters
with their higher-rank $T_{3}$ counterpart is the largest, it reduces the 10.971 and 14.636 millihartree
errors relative to CCSDT obtained with CR-CC(2,3) and the similarly large errors obtained with
CCSD(T) to a 0.1 millihartree level using as little as 2--3\% of all triply excited determinants
in the underlying $P$ spaces.
With only 2\% of all triples in the $P$ space, the difference between the activation energies characterizing the
automerization of cyclobutadiene obtained with the relaxed variant of the adaptive \ccpq{} approach and full
CCSDT is less than 0.1 kcal/mol, as opposed to the orders of magnitude larger, 8--9 kcal/mol, errors
relative to CCSDT resulting from the CR-CC(2,3) and CCSD(T) calculations. The rate with which the energies
resulting from the adaptive \ccpq{} calculations based on the relaxed algorithm approach the parent CCSDT
energetics, observed in Tables \ref{tab:table1} and \ref{tab:table3}, is certainly most encouraging.

As one might anticipate, and as confirmed in Tables \ref{tab:table1} and \ref{tab:table3}, the unrelaxed variant of
the adaptive \ccpq{}  methodology, in which one picks a particular fraction of triples for inclusion in the
$P$ space based on their contributions to the CR-CC(2,3) correction to CCSD, is less accurate for the stretched
\ftwoplus{} molecule, the TS structure of cyclobutadiene, and the associated barrier height than its relaxed
counterpart, which reaches a desired fraction of triply excited determinants in the $P$ space through a
sequence of recursively generated subspaces, but the results of the unrelaxed \ccpq{} calculations,
especially given their simplicity and lower computational costs, are excellent too. As shown in Tables
\ref{tab:table1} and \ref{tab:table3}, with only 2--3\% of all triply excited determinants
in the underlying $P$ spaces, the adaptive \ccpq{} computations based on the unrelaxed algorithm reduce the
9--11 millihartree, 14--15 millihartree, and 8--9 kcal/mol unsigned errors relative to CCSDT characterizing
the CCSD(T) and CR-CC(2,3) calculations for the $r =2 r_{e}$ \ftwoplus{} system, the TS species involved in
the automerization of cyclobutadiene, and the corresponding barrier height, respectively, to a chemical accuracy
(1 millihartree or 1 kcal/mol) level. Compared to the adaptive \ccpq{} calculations using the relaxed scheme,
the convergence rate toward the parent CCSDT energetics characterizing the unrelaxed approach is slower,
but the fact that one can obtain such high accuracies with just a few percent of all triples in the
underlying $P$ spaces, when the CCSD(T) and CR-CC(2,3) corrections to CCSD fail or struggle, is
encouraging. It is worth noticing that with only 1\% of all triply excited determinants in the
relevant
$P$ spaces,
which is the smallest fraction of triples considered
in this work, the adaptive \ccpq{} computations reduce the 10.971 millihartree, 14.636 millihartree, and
8.653 kcal/mol errors obtained with CR-CC(2,3)
for the challenging $r =2 r_{e}$ \ftwoplus{} system, the TS structure of cyclobutadiene,
and the activation energy characterizing the automerization of cyclobutadiene relative to CCSDT to
2.173 millihartree, 0.601 millihartree, and 0.412 kcal/mol, respectively. Given the fact that 1\% is also the incremental
fraction of triples used to enlarge the $\HP{k}$ spaces in the relaxed \ccpq{} calculations reported in
this study, the relaxed and unrelaxed \ccpq{} computations using the leading 1\% of all triply excited determinants 
identified by the adaptive \ccpq{} algorithm are equivalent.

While the $P$ spaces containing only 1\% of all triply excited determinants may not be rich enough to
bring the errors characterizing the adaptive \ccpq{} calculations for the stretched \ftwoplus{} ion and the
TS structure of cyclobutadiene relative to CCSDT to a 0.1 millihartree level, they are sufficient for
achieving high accuracies of this type in the adaptive \ccpq{} computations for the stretched fluorine
molecule and the cyclobutadiene R species. As shown in Tables \ref{tab:table1} and \ref{tab:table3},
the adaptive \ccpq{} calculations for the $r =2 r_{e}$ \ftwo{} system and the R structure of cyclobutadiene
using only 1\% of all triply excited determinants in the underlying $P$ spaces reduce the 4.254 and 0.848
millihartree differences between the CR-CC(2,3) and CCSDT energies and the $26.354$ and 1.123 millihartree unsigned
errors relative to CCSDT obtained with CCSD(T) to about 60 microhartree. Clearly, these are dramatic
improvements, especially given the small computational effort involved. Furthermore, unlike in the
$r =2 r_{e}$ \ftwoplus{} and cyclobutadiene TS systems, the results of the adaptive \ccpq{} computations
for the stretched fluorine molecule and the R species of cyclobutadiene using larger fractions of triples
in the corresponding $P$ spaces do not change much when the relaxed algorithm is replaced by its simpler unrelaxed
counterpart. We can rationalize these observations as follows. In the case of the stretched \ftwo{} molecule,
$T_{3}$ correlations are large and nonperturbative, so that one is much better off by using the CR-CC(2,3)
triples correction to CCSD instead of CCSD(T), but the coupling of $T_{1}$ and $T_{2}$ clusters with $T_{3}$ is not
as well pronounced as in the stretched \ftwoplus{} and TS species of cyclobutadiene and can, therefore,
be captured by injecting a tiny fraction of the leading triply excited determinants
into the \ccp{} calculations preceding the determination of the noniterative $\delta_{0}(P;Q)$ correction.
As a result, the incorporation of a larger fraction of triples in the underlying $P$ space and the
iterative construction of the $P$ space via the relaxed algorithm are not
necessary to obtain high accuracies in the adaptive \ccpq{} computations for the stretched
\ftwo{}. In the case of the R species of cyclobutadiene, $T_{3}$ correlations
are relatively small, perturbative, and largely decoupled from those captured by $T_{1}$ and $T_{2}$ clusters
and their powers entering the CC wave function ansatz, making the need for the use of larger fractions of
triples in the $P$ space and the iteratively constructed $\HP{k}$ spaces even smaller.

It is interesting to observe that many of the above comments also apply to the CCSD(T)-like variant
of the adaptive \ccpq{} methodology, abbreviated as \ccpqT{}, which is obtained by approximating
moments $\mathfrak{M}_{abc}^{ijk}(P^{(k)})$ and coefficients $\ell_{ijk}^{abc}(P^{(k)})$ in Eq.
(\ref{eq:delta_ijkabc}) such that the initial, $k=0$, \ccpqT{} energies are identical to those
obtained with CCSD(T). As shown in Tables \ref{tab:table2} and \ref{tab:table4}, the adaptive
\ccpqT{} computations using as little as 1--3\% of all triply excited determinants in the
underlying $P$ spaces, identified with the help of the relaxed or unrelaxed variants of the
\ccpqT{} algorithm, offer major error reductions relative to CCSDT compared to the CCSD(T) and
CR-CC(2,3) calculations. As one might anticipate, the results of the \ccpqT{}
calculations are not as accurate as those obtained with the more complete form of the adaptive
\ccpq{} formalism used in Tables \ref{tab:table1} and \ref{tab:table3}, in which the approximations in
moments $\mathfrak{M}_{abc}^{ijk}(P^{(k)})$ and coefficients $\ell_{ijk}^{abc}(P^{(k)})$ entering
Eq. (\ref{eq:delta_ijkabc}) that we invoked when implementing these quantities in our CCpy code are much
less drastic than in the \ccpqT{} case. Nevertheless, it is encouraging to observe that the replacement
of Eq. (\ref{eq:delta_ijkabc}) by its CCSD(T)-like counterpart does not result in an erratic behavior
characterizing the CCSD(T) calculations for the stretched \ftwo{} and \ftwoplus{} molecules and the
cyclobutadiene TS species, where electronic quasidegeneracies are substantial and $T_{3}$ correlations
become large and nonperturbative. The reduction of the large unsigned errors relative to CCSDT obtained with
CCSD(T) for the $r = 2r_{e}$ \ftwo{} and \ftwoplus{} systems, the TS structure of cyclobutadiene, and the
corresponding barrier height, of about 26, 9, and 14 millihartree and more than 8 kcal/mol, respectively,
to 1--2 millihartree and $<$1 kcal/mol levels by the adaptive \ccpqT{} approach using only 1\% of all
triples in the underlying $P$ spaces is certainly promising. Similarly encouraging are the major
improvements in the energies of the $r = 2r_{e}$ \ftwoplus{} system and the TS species of cyclobutadiene,
where one has to relax the lower-rank $T_{1}$ and $T_{2}$ amplitudes in the presence of their higher-rank
$T_{3}$ counterparts to obtain a reliable description, resulting from the replacement of the CR-CC(2,3)
triples correction to CCSD by the adaptive \ccpqT{} scheme using tiny fractions of triples in the underlying
$P$ spaces.

The fact that the adaptive \ccpqT{} approach works as well as it does is a consequence of including
the largest triply excited cluster amplitudes in the iterative \ccp{} steps preceding the determination
of the noniterative $\delta_{0}(P;Q)$ corrections [which take care of the remaining $T_{3}$ correlations
not included in the \ccp{} calculations]. The largest triply excited cluster amplitudes engage valence
shells that become quasidegenerate in multireference situations and cannot, as such, be accurately estimated
using the arguments originating from the many-body perturbation theory exploited in the design of the (T)
correction of CCSD(T) because of the smallness of the M{\o}ller--Plesset-type denominators. They are the ones
causing failures of the triples corrections of CCSD(T). After solving for the largest triply excited cluster
amplitudes using the iterative \ccp{} approach, one is left with the remaining, generally much smaller and
largely perturbative, $T_{3}$ contributions that can be reasonably well described by the CCSD(T)-like
expressions (cf. Refs.\ \onlinecite{rmrccsdt,h1} for similar considerations). One might argue that the adaptive
\ccpqT{} approach using a simplified form of the triples correction $\delta_{0}(P;Q)$, reminiscent of that
adopted by CCSD(T), which does not, for example, require solving the left eigenvalue problem involving
$\overline{H}^{(P)}(2)$ to construct the coefficients $\ell_{ijk}^{abc}(P^{(k)})$ entering Eq.
(\ref{eq:delta_ijkabc}), is a good substitute for its more complete \ccpq{} parent used in Tables \ref{tab:table1}
and \ref{tab:table3}, but we prefer the latter approach for the following two reasons. First, a direct comparison
of the results of the adaptive \ccpqT{} calculations reported in Tables \ref{tab:table2} and \ref{tab:table4}
with their \ccpq{} counterparts shown in Tables \ref{tab:table1} and \ref{tab:table3} demonstrates that the more
complete treatment of moments $\mathfrak{M}_{abc}^{ijk}(P^{(k)})$ and coefficients $\ell_{ijk}^{abc}(P^{(k)})$
in Eq. (\ref{eq:delta_ijkabc}), adopted in the latter scheme, results in generally higher accuracies. Second,
all of the computational timings that we have analyzed to date, including those presented in
Table \ref{tab:table5}, which we discuss next, show that savings in the computational effort offered
by the CCSD(T)-like \ccpqT{} scheme, compared to
its CR-CC(2,3)-like \ccpq{} counterpart, are not
large enough to favor the adaptive \ccpqT{} treatment over its slightly more expensive,
but also more robust, \ccpq{} parent.

\subsection{Performance of the adaptive CC(\protect\mbox{\boldmath$P$;$Q$}) and
CC(\protect\mbox{\boldmath$P$;$Q$})$_{\textbf{(T)}}$ approaches: Computational effort}
\label{sec3.3}

To illustrate the computational benefits of employing the adaptive \ccpq{} methodology as a substitute for full
CCSDT, which the adaptive \ccpq{} computations using tiny fractions of triples in the underlying $P$ spaces
approximate so well, in Table \ref{tab:table5} we show the timings characterizing the various CC calculations
carried out in this study for the challenging TS structure of cyclobutadiene, as described by the cc-pVDZ basis.
As in the case of Tables \ref{tab:table1}--\ref{tab:table4}, the information about the adaptive \ccpq{} and \ccpqT{}
calculations and the associated CCSD, CCSD(T), and CR-CC(2,3) runs included in Table \ref{tab:table5} was
obtained with our standalone CCpy package implemented in Python and available on GitHub,\cite{CCpy-GitHub}
which uses the Numpy library and Fortran routines in the computationally intensive parts, whereas the CPU time
needed to converge the reference CCSDT energy was determined using the highly efficient Fortran code developed
in the context of our CC(t;3) and other active-orbital-based \ccpq{} studies in Refs.\
 \onlinecite{jspp-chemphys2012,jspp-jcp2012,nbjspp-molphys2017}.
All of the CPU timings shown in Table \ref{tab:table5} correspond to single-core runs without taking advantage
of the $D_{4h}$ spatial symmetry of the TS structure of cyclobutadiene or its $D_{2h}$ Abelian subgroup
in the post-RHF steps. The computational times associated with the execution of the integral, SCF, and
integral transformation routines used to generate the one- and two-electron molecular integrals in the
RHF basis, which was carried out using GAMESS, and the integral sorting operations preceding the
CC steps performed using CCpy are ignored in Table \ref{tab:table5}.

To facilitate our discussion, in reporting the computational timings that characterize our adaptive \ccpq{}
and \ccpqT{} runs for the cyclobutadiene/cc-pVDZ system in its TS geometry, we focus on the calculations using
1\%, 3\%, and 5\% of the $S_z = 0$ triply excited determinants of the $\mathrm{A}_{1g} (D_{2h})$ symmetry
in the corresponding $P$ spaces identified by the unrelaxed variants of the adaptive \ccpq{} and \ccpqT{}
algorithms which, as shown in Tables \ref{tab:table3} and \ref{tab:table4}, reduce the 14.198 and 14.636
millihartree errors relative to CCSDT obtained with CCSD(T) and CR-CC(2,3) to as little as 0.601, 0.561, and
0.559 millihartree, respectively, in the case of \ccpq{}, and $-1.618$, $-0.685$, and $-0.347$ millihartree,
respectively, when the unrelaxed \ccpqT{} approach is employed. To get useful insights,
in addition to the total CPU times, we report the timings associated with the three key stages of the
adaptive \ccpq{} and \ccpqT{} calculations. In the case of the unrelaxed \ccpq{} and \ccpqT{} algorithms
considered in Table \ref{tab:table5}, these three key stages include
(i) the $P$ space determination, which consists of the initial CR-CC(2,3) or CCSD(T) run
followed by the analysis of the $\delta_{ijk,abc}(0)$ contributions to the resulting triples
correction to CCSD, needed to identify a desired fraction of triply excited determinants for inclusion
in the subsequent \ccp{} computation,
(ii) the iterative \ccp{} calculation using the $P$ space $\HP{1}$ consisting of all singly and doubly excited
determinants and a subset of triply excited determinants identified in stage (i),
and
(iii) the determination of the noniterative $\delta_{0}(P;Q)$ correction to the \ccp{} energy
obtained in stage (ii) to capture the remaining $T_{3}$ correlations with the help of the
complementary $Q$ space $\HQ{1}$ using Eq. (\ref{eq:delta_k}) in which we set $k$ at 1.

At this point, we can handle the preparatory stage (i), needed to filter out a particular fraction of the triply
excited determinants characterized by the largest $|\delta_{ijk,abc}(k)|$ [in the unrelaxed calculations
considered in Table \ref{tab:table5}, $|\delta_{ijk,abc}(0)|$]
values, in two different ways. In a faster, but also more memory intensive, scheme,
further elaborated on in footnote (e) of Table \ref{tab:table5}, we identify a desired fraction of the triply
excited determinants for inclusion in the $P$ space $\HP{k}$ [in the unrelaxed case, $\HP{1}$] after storing
all $\delta_{ijk,abc}(k)$ [in Table \ref{tab:table5}, $\delta_{ijk,abc}(0)$] contributions in memory. As shown
in Table \ref{tab:table5}, the CPU timings associated with this scheme are small, considerably smaller than the
computational times required by the \ccp{} iterations, and barely dependent on the
fraction of triples one is interested in identifying. In the second determinant selection scheme, which
is characterized by the minimum memory requirements that are as low as the fraction of triples one would
like to include in the iterative \ccp{} steps, described in further detail in footnote (f) of
Table \ref{tab:table5}, an array that has a dimension equal to the number of triples included in the
$P$ space of a given \ccpq{} or \ccpqT{} run is populated and continually repopulated with the
small subsets of $\delta_{ijk,abc}(k)$ [in the unrelaxed case, $\delta_{ijk,abc}(0)$] contributions as
they are being computed, analyzed, and filtered out to identify those with the largest $|\delta_{ijk,abc}(k)|$
values. The main advantage of this scheme, when applied to small fractions of triples included in the
relevant $P$ spaces, is the fact that its memory requirements are consistent with the small numbers of triples
included in the \ccp{} iterations. The problem with this scheme is a steep increase of the computational
time with the fraction of triples included in the $P$ space, which may, as shown in Table
\ref{tab:table5}, substantially exceed the time spent
on the remaining steps of the adaptive \ccpq{} and \ccpqT{} runs. This is not an issue when the fraction of
triples included in the $P$ space is 1--2\%, but it may become a problem when larger fractions must be considered.
This does not appear to be a major obstacle, since the convergence of the adaptive \ccpq{} calculations toward
CCSDT is fast and the results obtained with 1--2\% of triples in the $P$ space are already in
excellent agreement with the parent CCSDT data, greatly improving the CCSD(T) and CR-CC(2,3) energetics at
very small fractions of the costs of CCSDT computations. Nevertheless, we will continue working on our low-memory
determinant selection algorithm to reduce the CPU timings associated with it. One of the ideas that we plan
to experiment with is the use of somewhat larger arrays that are populated and repopulated with
subsets of $\delta_{ijk,abc}(k)$ contributions in our low-memory determinant selection scheme, which could,
for example, store as many $\delta_{ijk,abc}(k)$ values as a desired fraction of triples to be included in
the $P$ space multiplied by a small prefactor to reduce the operation count compared to the current
minimalistic implementation.
We will also investigate if one can reduce the number of the candidate $Q$-space determinants screened
by our low-memory selection algorithm, which might substantially improve its efficiency too. In particular,
we will examine if it is possible to adopt the determinant selection procedures analogous to those discussed
in Section III.D.2 of Ref. \onlinecite{cipsi_2}, used in CIPSI, or exclude the unimportant triply excited
determinants characterized by tiny $|\delta_{ijk,abc}(k)|$ values from the set of the $Q$-space triples
screened by the low-memory selection algorithm described in footnote (f) of Table \ref{tab:table5}.
Although the faster, but also much more memory intensive, determinant selection algorithm summarized in
footnote (e) of Table \ref{tab:table5} is quite efficient as is, as far as the operation count is concerned,
excluding the unimportant triply excited determinants from the set of the $Q$-space triples prior to filtering
out those that are characterized by the largest $|\delta_{ijk,abc}(k)|$ values will benefit this algorithm too,
speeding it up and reducing its memory requirements.

The illustrative CPU timings included in Table \ref{tab:table5}, combined with the previously discussed
energetics compared to CCSDT in Tables \ref{tab:table1}--\ref{tab:table4}, clearly demonstrate the enormous
benefits of using the adaptive \ccpq{} methodology proposed in this study. If we focus on the faster
determinant selection algorithm, as summarized above, the adaptive \ccpq{} runs using 1--5\% of triples
in the underlying $P$ spaces accelerate the parent CCSDT computations for the cyclobutadiene/cc-pVDZ system
in its TS geometry with only minimal loss of accuracy by factors on the order of 67--81, reducing the $>$14
millihartree errors relative to CCSDT obtained with the CCSD(T) and CR-CC(2,3) triples corrections to CCSD
to small fractions of a millihartree. The speedups offered by the adaptive \ccpq{} runs using the slower,
minimum-memory, $P$ space determination algorithm are not as good when the fractions of triples in the $P$
space are larger, but they are still quite impressive, reducing the CPU time required by the CCSDT calculation
for the cyclobutadiene/cc-pVDZ TS species by factors of 34 and 17 when 3\% and 5\% of all triples, respectively,
are included in the underlying $P$ spaces. We believe that we will be able to considerably improve these
speedups, while keeping the memory requirements associated with the $P$ space determination algorithm at
low levels. In the case of having 1\% of all triples in the $P$ space, the performance of our low-memory
$P$ space determination algorithm is excellent. The resulting adaptive \ccpq{} calculations for the
cyclobutadiene/cc-pVDZ system in its TS geometry are 75 times faster than the parent CCSDT run,
while reducing the $>$14 millihartree errors relative to CCSDT obtained in the CCSD(T) and CR-CC(2,3)
computations to about 0.6 millihartree.

The computational timings collected in Table \ref{tab:table5} allow us to draw a few other conclusions.
While we will continue working on improving our codes, the results in Table \ref{tab:table5} already show
a reasonably high efficiency of our current implementation of the adaptive \ccpq{} and \ccpqT{} approaches
incorporated in the CCpy package available on GitHub.\cite{CCpy-GitHub} This is particularly true in the
case of the most challenging \ccp{} routines which, in analogy to the previously formulated semi-stochastic
\cite{stochastic-ccpq-prl-2017,stochastic-ccpq-molphys-2020,stochastic-ccpq-jcp-2021,%
stochastic-ccpq-biradicals-jcp-2022} and CIPSI-driven\cite{cipsi-ccpq-2021} CC($P$;$Q$) approaches,
are used to solve the CC equations in the $P$ spaces containing spotty subsets of triply excited
determinants that do not form continuous manifolds labeled by occupied and unoccupied orbitals from
the respective ranges of indices assumed in conventional CC programming. We did discuss
some of the key algorithmic ingredients that must be considered when coding the \ccpq{} approaches,
if we are to benefit from the potentially enormous speedups offered by methods in this category when
the excitation manifolds used in the iterative \ccp{} steps [and their excited-state EOMCC($P$) counterparts]
are small and spotty, in Refs.\
\onlinecite{stochastic-ccpq-prl-2017,stochastic-ccpq-molphys-2020,stochastic-ccpq-jcp-2021}
(cf., also, Ref.\ \onlinecite{eomccp-jcp-2019}), but we will return to this topic in a separate publication,
where the coding strategies adopted in our current implementation of the adaptive \ccpq{} and \ccpqT{} methods
in the CCpy package will be discussed in detail. One of the most useful consequences of our current
implementation of the adaptive \ccpq{} and \ccpqT{} approaches aimed at accurately approximating the
full CCSDT energetics, in addition to the major savings in the computational effort compared to CCSDT,
is the observation that the CPU timings characterizing the \ccp{} calculations preceding the
determination of the noniterative $\delta_{0}(P,Q)$ corrections grow slowly with the fraction of triply excited
determinants included in the underlying $P$ spaces (see Table \ref{tab:table5}). This is consistent with
the computational cost analysis of the \ccpq{} methods using small fractions of higher--than--doubly excited
cluster and excitation amplitudes in the iterative \ccp{} and EOMCC($P$) steps in Refs.\
\onlinecite{stochastic-ccpq-prl-2017,eomccp-jcp-2019,stochastic-ccpq-molphys-2020,stochastic-ccpq-jcp-2021}.

Among other observations worth mentioning here, which are clearly reflected in the CPU timings shown in
in Table \ref{tab:table5} and the results of the adaptive \ccpq{} and \ccpqT{} calculations reported
in Tables \ref{tab:table1}--\ref{tab:table4}, and which are also consistent with our earlier active-orbital-based,
\cite{jspp-chemphys2012,jspp-jcp2012,jspp-jctc2012,nbjspp-molphys2017,ccpq-be2-jpca-2018,ccpq-mg2-mp-2019}
semi-stochastic,\cite{stochastic-ccpq-prl-2017,stochastic-ccpq-molphys-2020,stochastic-ccpq-jcp-2021,%
stochastic-ccpq-biradicals-jcp-2022} and
CIPSI-driven\cite{cipsi-ccpq-2021} CC($P$;$Q$) studies, is the fact that one is much better off by
adding the $\delta_{0}(P;Q)$ corrections to the \ccp{} energies. The uncorrected \ccp{} energies
approach their CCSDT parents as the number of triply excited determinants in the $P$ space
increases, improving the associated initial CCSD energetics and the CCSD(T) and CR-CC(2,3) results,
but the $\delta_{0}(P;Q)$-corrected \ccpq{} and \ccpqT{} energy values are much more accurate, converging
to CCSDT much faster, while the computational costs of determining the $\delta_{0}(P;Q)$ corrections are
only small fractions of those associated with the underlying \ccp{} steps. We must also realize
that we have to determine the $\delta_{0}(P;Q)$ [in general, $\delta_{\mu}(P;Q)$] corrections in
the adaptive \ccpq{} and \ccpqT{} schemes anyway, since we need them to construct the underlying $P$ spaces,
so using the uncorrected \ccp{} [in the case of excited states, EOMCC($P$)] energies does not make much sense
and is not recommended. This is, in a way, similar to the selected CI calculations, such as the CIPSI approach used
as an inspiration in the development of the adaptive \ccpq{} framework in this study, which rely on the perturbative 
corrections to the variational CI energies in the process of constructing the underlying Hamiltonian
diagonalization spaces. It is also clear from the timings reported in Table \ref{tab:table5} and
comparisons with the CCSDT energetics in Tables \ref{tab:table1}--\ref{tab:table4} that there are
no major benefits of using the CCSD(T)-like \ccpqT{} approximation to the adaptive \ccpq{} methodology,
which relies on a much more complete and robust treatment of moments $\mathfrak{M}_{abc}^{ijk}(P^{(k)})$
and coefficients $\ell_{ijk}^{abc}(P^{(k)})$ in Eq. (\ref{eq:delta_ijkabc}) compared to that employed in \ccpqT{}.
The adaptive \ccpqT{} energies are generally less accurate, while the savings in the computational effort
offered by the \ccpqT{} approach compared to its more complete, CR-CC(2,3)-like, \ccpq{} counterpart are small.

Last, but not least, it is interesting to observe in Table \ref{tab:table5} that the CPU timings
characterizing the adaptive \ccpq{} computations capable of providing the near-CCSDT energetics,
while larger than those required by their CCSD(T) and CR-CC(2,3) counterparts, are not very far removed from the
costs of the CCSD(T) and CR-CC(2,3) calculations, which are generally less robust and much less accurate,
especially when $T_{3}$ correlations are larger and nonperturbative and the coupling between
the lower-rank $T_{1}$ and $T_{2}$ and higher-rank $T_{3}$ clusters becomes significant. This is
mainly because the CPU times needed to complete the iterative \ccp{} steps, when the fraction of
triples incorporated in the $P$ space is as tiny as 1-3\%, which is, as shown in Tables
\ref{tab:table1}--\ref{tab:table4}, enough to recover the CCSDT energetics to within
fractions of a millihartree, even when CCSD(T) and CR-CC(2,3) fail or struggle, are only a few times
longer than those characterizing CCSD. This is not too surprising. When the fraction of triples in the $P$
space is tiny, the noniterative triples corrections of the adaptive \ccpq{} and CR-CC(2,3) calculations have
virtually identical costs, since the corresponding $Q$ spaces, which are spanned by either all [CR-CC(2,3)] or
nearly all [\ccpq{}] triply excited determinants, are very similar. One can, therefore, anticipate
that much of the difference between the CPU timings characterizing the CR-CC(2,3)
and adaptive \ccpq{} runs, especially if we put aside the $P$ space determination steps, is
driven by the time spent on the \ccp{} iterations, and this is exactly what we see in Table \ref{tab:table5}.
Similar remarks apply to the relative costs of the CCSD(T) vs. adaptive \ccpqT{} computations. The storage
requirements characterizing the adaptive \ccpq{} and \ccpqT{} calculations are not far removed from those
required by their CCSD(T) and CR-CC(2,3) counterparts either, if the fractions of triples in the underlying
$P$ spaces are tiny, since the vectors representing the $T^{(P)} = T_{1} + T_{2} + T_{3}^{(P)}$
operators are not much longer than those used to store $T_{1} + T_{2}$. In summary, it is enouraging that
with the introduction of the adaptive \ccpq{} methodology, which is only a few times more expensive than
the CCSD(T) and CR-CC(2,3) approaches, we can accelerate the full CCSDT computations by orders of
magnitude with only minimal loss of accuracy, even when $T_{3}$ effects are highly nonperturbative, causing
CCSD(T) to fail, and even when the coupling among $T_{1}$, $T_{2}$, and $T_{3}$ components of the cluster
operator becomes so large that the CR-CC(2,3) correction to CCSD is no longer reliable.

\section{Summary, Concluding Remarks, and Future Outlook}
\label{sec4}

One of the most important topics in the CC theory and its EOM and linear-response extensions to excited
electronic states has been the development of accurate and robust ways of incorporating higher--than--two-body
components of the cluster and excitation operators that would avoid large, usually prohibitive, computational
costs of methods such as CCSDT, CCSDTQ, EOMCCSDT, etc., which treat these operators fully, while eliminating
failures of the more practical perturbative approximations of the CCSD(T) type in typical multireference situations,
such as chemical bond stretching or breaking, biradicals, and excited states dominated by two-electron
transitions. Among the most promising ideas in this area has been the CC($P$;$Q$) formalism,
\cite{jspp-chemphys2012,jspp-jcp2012,jspp-jctc2012,nbjspp-molphys2017,ccpq-be2-jpca-2018,ccpq-mg2-mp-2019,%
stochastic-ccpq-prl-2017,stochastic-ccpq-molphys-2020,stochastic-ccpq-jcp-2021,stochastic-ccpq-biradicals-jcp-2022,%
cipsi-ccpq-2021} in which one solves the CC/EOMCC equations in a subspace of the many-electron Hilbert space,
 called the $P$ space, spanned by the excited determinants that, together with the reference determinant,
dominate the ground or the ground  and excited states of interest, and then improves the resulting \ccp{}
or \ccp{} and EOMCC($P$) energies using the {\it a posteriori} noniterative $\delta_{\mu}(P;Q)$ corrections due
to higher-order correlation effects not captured by the CC($P$)/EOMCC($P$) computations, determined with
the help of another subspace of the Hilbert space, called the $Q$ space. The CC($P$;$Q$) methodology
is
a generalization of the biorthogonal moment expansions of Refs.\ \onlinecite{crccl_jcp,crccl_cpl,%
crccl_molphys,crccl_ijqc,crccl_ijqc2}, which have resulted in methods such as CR-CC(2,3), but unlike
its predecessors, it permits unconventional choices of the $P$ and $Q$ spaces that
relax the lower-rank $T_{n}$ and $R_{\mu,n}$
components in the presence
of their higher-rank counterparts,
which CR-CC(2,3), CCSD(T), and
their higher-order and excited-state extensions
cannot do.

In this work,
we have introduced a novel category
of the \ccpq{} formalism, called adaptive \ccpq{}, which applies to ground and excited states and which results
in methods that are expected to converge or accurately approximate high-level CC/EOMCC energetics of the CCSDT,
CCSDTQ, EOMCCSDT, etc. types at small fractions of the computational costs. The adaptive \ccpq{} approaches should
work equally well in cases of weaker as well as stronger correlations, from nondegenerate ground states and
excited states dominated by one-electron transitions to bond stretching or breaking, biradicals,
and excited states having substantial double excitation character, where higher--than--two-body
components of the cluster and EOM excitation operators become large and can be strongly coupled to their
lower-rank counterparts. The adaptive \ccpq{} methodology discussed in this
paper is a ``black-box'' and self-improving framework. The key idea behind its design is an
adaptive selection of the leading determinants or excitation amplitude types needed to define the $T_{n}$
(in the case of excited states, also $R_{\mu,n}$) components with $n > 2$ that enter the
CC($P$) and EOMCC($P$) calculations, which relies on the intrinsic mathematical structure of the moment
expansions defining the noniterative $\delta_{\mu}(P;Q)$ corrections to the CC($P$) and EOMCC($P$)
energies. This makes the process of identifying the leading higher--than--doubly excited determinants
for inclusion in the $P$ space fully automated. Prior to this work, in our quest to find an efficient and
robust procedure for selecting the appropriate subset of higher--than--doubly excited determinants for
inclusion in the $P$ space, we have relied on active orbitals,
\cite{jspp-chemphys2012,jspp-jcp2012,jspp-jctc2012,nbjspp-molphys2017,ccpq-be2-jpca-2018,ccpq-mg2-mp-2019}
CIQMC and CCMC wave function propagations,
\cite{stochastic-ccpq-prl-2017,eomccp-jcp-2019,stochastic-ccpq-molphys-2020,stochastic-ccpq-jcp-2021,%
stochastic-ccpq-biradicals-jcp-2022}
and sequences of Hamiltonian diagonalizations employed in the modern formulation of CIPSI,\cite{cipsi-ccpq-2021}
and all of these efforts have been successful, allowing us to recover or accurately approximate the
CCSDT,\cite{jspp-chemphys2012,jspp-jcp2012,jspp-jctc2012,nbjspp-molphys2017,ccpq-be2-jpca-2018,ccpq-mg2-mp-2019,%
stochastic-ccpq-prl-2017,stochastic-ccpq-molphys-2020,stochastic-ccpq-jcp-2021,cipsi-ccpq-2021,%
stochastic-ccpq-biradicals-jcp-2022}
CCSDTQ,\cite{nbjspp-molphys2017,ccpq-be2-jpca-2018,ccpq-mg2-mp-2019,stochastic-ccpq-jcp-2021}
and EOMCCSDT \cite{stochastic-ccpq-molphys-2020} energetics at small fractions of the computational costs,
but the adaptive CC($P$;$Q$) formalism, introduced and tested in this study, by eliminating
the need for reliance on user-defined active orbitals, non-CC (CIQMC, CIPSI) wave functions, and
stochastic (CIQMC, CCMC) concepts, moves us to an entirely new level.

In defining the adaptive \ccpq{} methodology, especially its relaxed variant, we have been inspired by
the modern implementation of the CIPSI approach in Quantum Package 2.0, where one grows the
relevant, recursively generated,
Hamiltonian diagonalization spaces by arranging the candidate determinants from outside the
current space according to the magnitude of their contributions to perturbative corrections to the
Hamiltonian eigenvalues, using those characterized by the largest contributions to enlarge the
variational space.
Thus, we
define our adaptive \ccpq{} algorithm by
interpreting the \ccpq{} correction $\delta_{\mu}(P;Q)$, Eq.\ (\ref{eq:mmcorrection}), as a sum
of contributions due to the individual determinants from the $Q$ space. We then use the $Q$-space determinants
characterized by the largest (in absolute value) contributions to $\delta_{\mu}(P;Q)$ to enlarge the
current $P$ space, repeating the process, if need be, to generate the CIPSI-style sequence of the
slowly growing $P$ spaces and slowly shrinking $Q$ spaces adding up to the excitation
manifold of the parent high-level CC/EOMCC theory of interest.

To illustrate the benefits
of the adaptive \ccpq{}
framework, we have developed the adaptive \ccpq{} code targeting
the ground-state CCSDT energetics, which we incorporated in our open-source
CCpy package available on GitHub.\cite{CCpy-GitHub} The initial $P$ space
in this code
is spanned by singly and doubly excited determinants
and the initial $Q$ space consists of all triples, so that our sequence of \ccpq{} calculations starts
from CR-CC(2,3), and then we follow the above recursive procedure by transferring more and more triply 
excited determinants from the $Q$ to $P$ spaces. In the adaptive \ccpq{} computations reported
in this article, we have distinguished between the relaxed algorithm, in which one reaches a desired
fraction of triply excited determinants in the $P$ space through a sequence of systematically grown
smaller subspaces, relaxing the one-, two-, and three-body components of the underlying cluster operator
accordingly, and a one-shot unrelaxed approach, in which one picks a target fraction of triples for
inclusion in the $P$ space based on their contributions to
the CR-CC(2,3)
correction to CCSD.
We have also considered the relaxed and unrelaxed variants of the adaptive \ccpqT{} approach,
in which the \ccpq{} correction to the CC($P$) energy is calculated using the CCSD(T)-like formulas.

The relaxed and unrelaxed variants of the adaptive \ccpq{} and \ccpqT{} methods aimed at converging
the ground-state CCSDT energetics have been tested using the significantly stretched \ftwo{} and \ftwoplus{}
molecules, in which $T_{3}$ correlations are large and highly nonperturbative, causing CCSD(T) to fail,
and the reactant and transition-state structures involved in the automerization of cyclobutadiene. In
analogy to the stretched \ftwoplus{} system, the transition-state structure of cyclobutadiene is characterized
by large and nonperturbative $T_{3}$ effects, which are also strongly coupled to those
captured by $T_{1}$ and $T_{2}$ and their powers, so that both CCSD(T) and CR-CC(2,3) fail in these two cases.
The CCSD(T) and CR-CC(2,3) methods work well for the cyclobutadiene molecule in its reactant,
i.e., equilibrium, geometry,
but some nonnegligible errors relative to CCSDT, which has been treated in our numerical work
as a reference, remain. Our
calculations show that independent of the system considered in the calculations, the adaptive \ccpq{}
scheme using the relaxed algorithm, in which the numbers of triples in the relevant $P$ spaces are
increased using tiny 1\% increments, recovers the target CCSDT energetics to within 0.1 millihartree or so
using only 2--3 \% of all triples in the final $P$ spaces, i.e., after 3--4 cycles of the algorithm.
This applies to all four systems included in our tests, including the most challenging stretched \ftwoplus{}
species and the transition state of cyclobutadiene, and the activation barrier associated with it,
where both CCSD(T) and CR-CC(2,3) fail. The simpler unrelaxed variant of the adaptive \ccpq{} approach is
less accurate and displays slower convergence toward CCSDT, but with similar fractions of triples in the underlying
$P$ spaces, it has no problem with reaching chemical accuracy levels relative to CCSDT. Even with a tiny
1\% of all triply excited determinants in the $P$ spaces, the adaptive \ccpq{} algorithm
proposed in this study works excellent, recovering the CCSDT energetics to within 1--2 millihartree, even
when the errors in the CCSD(T) and CR-CC(2,3) data relative to CCSDT reach $\sim$10--30 millihartree
and the CR-CC(2,3) method
struggles. Our current implementation of the relaxed and unrelaxed
adaptive \ccpq{} schemes aimed
at CCSDT is efficient
enough to allow us to recover
the CCSDT energetics with the above accuracy levels using computational times that are orders of
magnitude smaller than those characterizing CCSDT and that are not very far removed from those characterizing
the corresponding CCSD(T) and CR-CC(2,3) calculations. While the \ccpqT{} approximation to the adaptive
\ccpq{} works well too, one is better off using its more complete \ccpq{} counterpart, since the
adaptive \ccpqT{} energies are generally less accurate, while the savings in the computational effort
offered by \ccpqT{} are
comparatively
small.

The results obtained in this study are most encouraging, but, given the novelty of the adaptive \ccpq{}
methodology proposed in this work and the limited numerical evidence to date, we have to examine greater
numbers of molecular examples and a wider range of situations, especially those where $T_{3}$ and other
higher--than--two-body clusters become large and difficult to converge, and extend our current adaptive
\ccpq{} codes to higher levels of the CC theory, such as CCSDTQ. In analogy to our semi-stochastic \ccpq{}
efforts,\cite{stochastic-ccpq-molphys-2020} we will also implement and test the excited-state extensions of
the adaptive \ccpq{} approaches, starting with those aimed at the EOMCCSDT energetics. The present article
has already demonstrated that the adaptive \ccpq{} formalism applies to ground as well as excited states and
that extending the adaptive \ccpq{} codes to excited states within the EOMCC framework is
feasible. Ultimately, we will examine if it is practical
to extend the adaptive \ccpq{} algorithm discussed in this
article to converge the full CC/EOMCC, meaning full CI, data, using the $Q$ spaces which are orthogonal
complements to the $P$ spaces used in the \ccp{}/EOMCC($P$) and \ccpq{} considerations.
In all of the relaxed \ccpq{} calculations reported in this article, we have assumed a 1\%
growth rate in the fractions of triples added to the underlying $P$ spaces,
but we are planning to explore other (smaller as well as larger) incremental fractions of triples
used to grow $P$ spaces in the relaxed variant of the adaptive \ccpq{} algorithm as well, to see if they can
substantially affect any of the conclusions of this
study. All
of the above plans will be accompanied by
improvements in
the adaptive \ccpq{} code incorporated in the CCpy software package
available on GitHub,\cite{CCpy-GitHub} especially  the low-memory variant of the determinant selection algorithm
needed to
filter out the
leading triply (in the future also quadruply) excited determinants
for inclusion in the $P$ spaces used in the adaptive \ccpq{} computations.

\begin{acknowledgments}
This work has been supported by the Chemical Sciences, Geosciences
and Biosciences Division, Office of Basic Energy Sciences, Office
of Science, U.S. Department of Energy (Grant No. DE-FG02-01ER15228 to P.P.).
\end{acknowledgments}

\section*{Data Availability}
The data that support the findings of this study are available within the article.


\section*{References}
\bibliographystyle{aipnum4-1} 
\renewcommand{\baselinestretch}{1.1}
%

\newpage

\clearpage


\onecolumngrid



\squeezetable
\begin{table*}[h!]
\caption{\label{tab:table1}%
Convergence of the energies resulting from the relaxed (Rel.) and unrelaxed (Unrel.)
variants of the adaptive \ccpq{} approach and the underlying \ccp{} computations
toward CCSDT for the \ftwo{} and \ftwoplus{} molecules described by
the cc-pVTZ basis set, in which the F--F bond lengths $r$ were fixed at
$2r_e$, with $r_{e}$ representing the relevant equilibrium geometries
(2.66816 bohr for \ftwo{} and 2.49822 bohr for \ftwoplus{}). The
\%T values are the percentages of the $S_z = 0$ triply excited determinants
of the $\mathrm{A}_{1g} (D_{2h})$, for \ftwo{}, and $\mathrm{B}_{3g} (D_{2h})$,
for \ftwoplus{}, symmetries identified by the adaptive CC($P$;$Q$) algorithm and
included, alongside all singles and doubles, in the respective $P$ spaces.
The $Q$ spaces used in computing the \ccpq{} corrections were defined as
the remaining triples not included in the associated $P$ spaces. In
increasing the numbers of triply excited determinants in the $P$ spaces employed
in the relaxed calculations, a 1\% growth rate was assumed throughout. In all
post-RHF (\ftwo{}) and post-ROHF (\ftwoplus{}) calculations, the two lowest
core orbitals were kept frozen.}
\footnotesize
\begin{ruledtabular}
\begin{tabular}{ccccccccc}
& \multicolumn{4}{c}{${\rm F}_2$} & \multicolumn{4}{c}{${\rm F}_2^+$} \\ 
\cline{2-5} \cline{6-9}
\multicolumn{1}{c}{\textrm{\%T}}& \multicolumn{1}{c}{\textrm{\ccp{}}} & \multicolumn{1}{c}{\textrm{\ccpq{}}} & 
\multicolumn{1}{c}{\textrm{\ccp{}}} & \multicolumn{1}{c}{\textrm{\ccpq{}}} &
\multicolumn{1}{c}{\textrm{\ccp{}}} & \multicolumn{1}{c}{\textrm{\ccpq{}}} &
\multicolumn{1}{c}{\textrm{\ccp{}}} & \multicolumn{1}{c}{\textrm{\ccpq{}}} \\ 
& \multicolumn{1}{c}{Unrel.\footnotemark[1]} & \multicolumn{1}{c}{Unrel.\footnotemark[1]} & 
\multicolumn{1}{c}{Rel.\footnotemark[1]} & \multicolumn{1}{c}{Rel.\footnotemark[1]} &
\multicolumn{1}{c}{Unrel.\footnotemark[1]} & \multicolumn{1}{c}{Unrel.\footnotemark[1]} &
\multicolumn{1}{c}{Rel.\footnotemark[1]} & \multicolumn{1}{c}{Rel.\footnotemark[1]} \\ 
\colrule
0 & 62.819\footnotemark[2] & 4.254\footnotemark[3] & 62.819\footnotemark[2] & 4.254\footnotemark[3]
  & 76.291\footnotemark[2] & 10.971\footnotemark[3] & 76.291\footnotemark[2] & 10.971\footnotemark[3] \\
1\footnotemark[4] & 3.076 & 0.063 & 3.076 & 0.063 & 6.071 & 2.173 & 6.071 & 2.173 \\
2 & 2.103 & 0.089 & 2.052 & 0.057 & 4.061 & 1.560 & 2.599 & 0.191 \\
3 & 1.586 & 0.104 & 1.539 & 0.070 & 2.970 & 1.146 & 1.707 & $-0.026$ \\
4 & 1.243 & 0.098 & 1.212 & 0.071 & 2.100 & 0.684 & 1.330 & $-0.007$ \\
5 & 1.009 & 0.105 & 0.985 & 0.080 & 1.680 & 0.549 & 1.077 & 0.010 \\
\cline{2-5} \cline{6-9} 
100 & \multicolumn{4}{c}{$-199.238344$\footnotemark[5]} & \multicolumn{4}{c}{$-198.606409$\footnotemark[5]}
\end{tabular}
\end{ruledtabular}
\footnotetext[1]{The \ccp{} and \ccpq{} energies are reported as errors relative to CCSDT in millihartree.}
\footnotetext[2]{Equivalent to CCSD.}
\footnotetext[3]{Equivalent to CR-CC(2,3).}
\footnotetext[4]{For \%T = 1, the \ccp{} and \ccpq{} energies obtained in 
the relaxed and unrelaxed calculations are identical.}
\footnotetext[5]{Total CCSDT energy in hartree.}
\end{table*}


\squeezetable
\begin{table*}[h!]
\caption{\label{tab:table2}%
Same as Table \ref{tab:table1} except that the \ccpq{} corrections to the \ccp{}
energies are replaced by their CCSD(T)-like \ccpqT{} counterparts. As in
Table \ref{tab:table1}, the F--F bond lengths $r$ in \ftwo{} and \ftwoplus{}
were fixed at $2r_e$, where $r_{e}$ designates the corresponding equilibrium geometries
(2.66816 bohr for \ftwo{} and 2.49822 bohr for \ftwoplus{}).}
\footnotesize
\begin{ruledtabular}
\begin{tabular}{ccccccccc}
& \multicolumn{4}{c}{${\rm F}_2$} & \multicolumn{4}{c}{${\rm F}_2^+$} \\ 
\cline{2-5} \cline{6-9}
\multicolumn{1}{c}{\textrm{\%T}} & \multicolumn{1}{c}{\textrm{\ccp{}}} & \multicolumn{1}{c}{\textrm{\ccpqT{}}} & 
\multicolumn{1}{c}{\textrm{\ccp{}}} & \multicolumn{1}{c}{\textrm{\ccpqT{}}} &
\multicolumn{1}{c}{\textrm{\ccp{}}} & \multicolumn{1}{c}{\textrm{\ccpqT{}}} &
\multicolumn{1}{c}{\textrm{\ccp{}}} & \multicolumn{1}{c}{\textrm{\ccpqT{}}} \\ 
& \multicolumn{1}{c}{Unrel.\footnotemark[1]} & \multicolumn{1}{c}{Unrel.\footnotemark[1]} & 
\multicolumn{1}{c}{Rel.\footnotemark[1]} & \multicolumn{1}{c}{Rel.\footnotemark[1]} &
\multicolumn{1}{c}{Unrel.\footnotemark[1]} & \multicolumn{1}{c}{Unrel.\footnotemark[1]} &
\multicolumn{1}{c}{Rel.\footnotemark[1]} & \multicolumn{1}{c}{Rel.\footnotemark[1]} \\ 
\colrule
0 & 62.819\footnotemark[2] & $-26.354$\footnotemark[3] & 62.819\footnotemark[2] & $-26.354$\footnotemark[3]
  & 76.291\footnotemark[2] & $-8.985$\footnotemark[3] & 76.291\footnotemark[2] & $-8.985$\footnotemark[3] \\
1\footnotemark[4] & 3.274 & $-0.789$ & 3.274 & $-0.789$ & 7.758 & 1.802 & 7.758 & 1.802 \\
2 & 2.205 & $-0.471$ & 2.316 & $-0.514$ & 5.861 & 1.991 & 3.513 & 0.010 \\
3 & 1.669 & $-0.309$ & 1.593 & $-0.349$ & 4.740 & 1.892 & 1.971 & $-0.445$ \\
4 & 1.314 & $-0.205$ & 1.253 & $-0.233$ & 3.930 & 1.697 & 1.478 & $-0.367$ \\
5 & 1.061 & $-0.145$ & 1.009 & $-0.166$ & 3.285 & 1.466 & 1.200 & $-0.264$ \\
\cline{2-5} \cline{6-9} 
100 & \multicolumn{4}{c}{$-199.238344$\footnotemark[5]} & \multicolumn{4}{c}{$-198.606409$\footnotemark[5]}
\end{tabular}
\end{ruledtabular}
\footnotetext[1]{The \ccp{} and \ccpqT{} energies are reported as errors relative to CCSDT in millihartree.}
\footnotetext[2]{Equivalent to CCSD.}
\footnotetext[3]{Equivalent to CCSD(T).}
\footnotetext[4]{For \%T = 1, the \ccp{} and \ccpqT{} energies obtained in 
the relaxed and unrelaxed calculations are identical.}
\footnotetext[5]{Total CCSDT energy in hartree.}
\end{table*}

\newpage


\squeezetable
\begin{table*}[h!]
\caption{\label{tab:table3}%
Convergence of the energies resulting from the relaxed (Rel.) and unrelaxed (Unrel.)
variants of the adaptive \ccpq{} approach and the underlying \ccp{} computations
toward CCSDT for the reactant (R) and transition-state (TS) structures involved in
the automerization of cyclobutadiene, as described by the cc-pVDZ basis set, optimized
in the MR-AQCC calculations reported in Ref.\ \onlinecite{MR-AQCC}, along with the
corresponding barrier heights. The \%T values are the percentages of the $S_z = 0$
triply excited determinants of the $\mathrm{A}_{1g} (D_{2h})$ symmetry identified by
the adaptive CC($P$;$Q$) algorithm and included, alongside all singles and doubles, in
the respective $P$ spaces. In analogy to \ftwo{} and \ftwoplus{}, the $Q$ spaces adopted
in computing the \ccpq{} corrections consisted of the triply excited determinants not
included in the associated $P$ spaces and in increasing the numbers of triples in the
$P$ spaces used in the relaxed calculations, a 1\% growth rate was assumed throughout.
In all post-RHF calculations, the four lowest core orbitals were kept frozen.}
\footnotesize
\begin{ruledtabular}
\begin{tabular}{ccccccccccc}
& \multicolumn{4}{c}{\textrm{R}} & \multicolumn{4}{c}{\textrm{TS}} & \multicolumn{2}{c}{\textrm{Barrier Height}} \\ 
\cline{2-5} \cline{6-9} \cline{10-11}
\multicolumn{1}{c}{\textrm{\%T}}& \multicolumn{1}{c}{\textrm{\ccp{}}} & \multicolumn{1}{c}{\textrm{\ccpq{}}} & 
\multicolumn{1}{c}{\textrm{\ccp{}}} & \multicolumn{1}{c}{\textrm{\ccpq{}}} &
\multicolumn{1}{c}{\textrm{\ccp{}}} & \multicolumn{1}{c}{\textrm{\ccpq{}}} &
\multicolumn{1}{c}{\textrm{\ccp{}}} & \multicolumn{1}{c}{\textrm{\ccpq{}}} &
\multicolumn{1}{c}{\textrm{\ccpq{}}} & \multicolumn{1}{c}{\textrm{\ccpq{}}} \\ 
& \multicolumn{1}{c}{Unrel.\footnotemark[1]} & \multicolumn{1}{c}{Unrel.\footnotemark[1]} & 
\multicolumn{1}{c}{Rel.\footnotemark[1]} & \multicolumn{1}{c}{Rel.\footnotemark[1]} &
\multicolumn{1}{c}{Unrel.\footnotemark[1]} & \multicolumn{1}{c}{Unrel.\footnotemark[1]} &
\multicolumn{1}{c}{Rel.\footnotemark[1]} & \multicolumn{1}{c}{Rel.\footnotemark[1]} & 
\multicolumn{1}{c}{Unrel.\footnotemark[2]} & \multicolumn{1}{c}{Rel.\footnotemark[2]} \\ 
\colrule
0 & 26.827\footnotemark[3] & 0.848\footnotemark[4] & 26.827\footnotemark[3] & 0.848\footnotemark[4]
  & 47.979\footnotemark[3] & 14.636\footnotemark[4] & 47.979\footnotemark[3] & 14.636\footnotemark[4]
  & 8.653\footnotemark[4] & 8.653\footnotemark[4] \\
1\footnotemark[5] & 12.622 & $-0.055$ & 12.622 & $-0.055$ & 12.622 & 0.601 & 12.622 & 0.601 & 0.412 & 0.412 \\
2 & 10.143 & $-0.013$ & 10.121 & $-0.030$ & 10.180 & 0.598 & 9.250 & $-0.169$ & 0.384 & $-0.087$ \\
3 & 8.610 & 0.016 & 8.588 & $-0.002$ & 8.643 & 0.561 & 7.821 & $-0.111$ & 0.343 & $-0.068$\\
4 & 7.501 & 0.037 & 7.482 & 0.022 & 7.571 & 0.568 & 6.827 & $-0.040$ & 0.334 & $-0.038$ \\
5 & 6.637 & 0.050 & 6.620 & 0.035 & 6.722 & 0.559 & 6.052 & 0.010 & 0.320 & $-0.016$ \\
\cline{2-5} \cline{6-9} \cline{10-11} 
100 & \multicolumn{4}{c}{$-154.244157$\footnotemark[6]} & \multicolumn{4}{c}{$-154.232002$\footnotemark[6]}
    & \multicolumn{2}{c}{7.627\footnotemark[7]}
\end{tabular}
\end{ruledtabular}
\footnotetext[1]{The \ccp{} and \ccpq{} energies of the R and TS species are reported
as errors relative to CCSDT in millihartree.}
\footnotetext[2]{The \ccpq{} values of the barrier height are reported as errors
relative to CCSDT in kcal/mol.}
\footnotetext[3]{Equivalent to CCSD.}
\footnotetext[4]{Equivalent to CR-CC(2,3).}
\footnotetext[5]{For \%T = 1, the \ccp{} and \ccpq{} energies obtained in
the relaxed and unrelaxed calculations are identical.}
\footnotetext[6]{Total CCSDT energy in hartree.}
\footnotetext[7]{The CCSDT barrier height in kcal/mol.}
\end{table*}


\squeezetable
\begin{table*}[h!]
\caption{\label{tab:table4}%
Same as Table \ref{tab:table3} except that the \ccpq{} corrections to the \ccp{}
energies are replaced by their CCSD(T)-like \ccpqT{} counterparts.
As in Table \ref{tab:table3}, the reactant (R) and transition-state (TS) structures
of cyclobutadiene adopted in the calculations, optimized using the MR-AQCC approach,
were taken from Ref.\ \onlinecite{MR-AQCC}.}
\footnotesize
\begin{ruledtabular}
\begin{tabular}{ccccccccccc}
& \multicolumn{4}{c}{\textrm{R}} & \multicolumn{4}{c}{\textrm{TS}} & \multicolumn{2}{c}{\textrm{Barrier Height}} \\ 
\cline{2-5} \cline{6-9} \cline{10-11}
\multicolumn{1}{c}{\textrm{\%T}} & \multicolumn{1}{c}{\textrm{\ccp{}}} & \multicolumn{1}{c}{\textrm{\ccpqT{}}} & 
\multicolumn{1}{c}{\textrm{\ccp{}}} & \multicolumn{1}{c}{\textrm{\ccpqT{}}} &
\multicolumn{1}{c}{\textrm{\ccp{}}} & \multicolumn{1}{c}{\textrm{\ccpqT{}}} &
\multicolumn{1}{c}{\textrm{\ccp{}}} & \multicolumn{1}{c}{\textrm{\ccpqT{}}} &
\multicolumn{1}{c}{\textrm{\ccpqT{}}} & \multicolumn{1}{c}{\textrm{\ccpqT{}}} \\ 
& \multicolumn{1}{c}{Unrel.\footnotemark[1]} & \multicolumn{1}{c}{Unrel.\footnotemark[1]} & 
\multicolumn{1}{c}{Rel.\footnotemark[1]} & \multicolumn{1}{c}{Rel.\footnotemark[1]} &
\multicolumn{1}{c}{Unrel.\footnotemark[1]} & \multicolumn{1}{c}{Unrel.\footnotemark[1]} &
\multicolumn{1}{c}{Rel.\footnotemark[1]} & \multicolumn{1}{c}{Rel.\footnotemark[1]} & 
\multicolumn{1}{c}{Unrel.\footnotemark[2]} & \multicolumn{1}{c}{Rel.\footnotemark[2]} \\ 
\colrule
0 & 26.827\footnotemark[3] & 1.123\footnotemark[4] & 26.827\footnotemark[3] & 1.123\footnotemark[4]
  & 47.979\footnotemark[3] & 14.198\footnotemark[4] & 47.979\footnotemark[3] & 14.198\footnotemark[4]
  & 8.205\footnotemark[4] & 8.205\footnotemark[4] \\
1\footnotemark[5] & 12.650 & $-0.347$ & 12.650 & $-0.347$ & 12.829 & $-1.618$ & 12.829 & $-1.618$
  & $-0.798$ & $-0.798$ \\
2 & 10.171 & $-0.254$ & 10.131 & $-0.263$ & 10.394 & $-1.009$ & 9.138 & $-1.417$ & $-0.473$ & $-0.723$ \\
3 & 8.638 & $-0.193$ & 8.603 & $-0.202$ & 8.894 & $-0.685$ & 7.729 & $-1.126$ & $-0.309$ & $-0.579$\\
4 & 7.530 & $-0.147$ & 7.498 & $-0.156$ & 7.802 & $-0.491$ & 6.746 & $-0.905$ & $-0.216$ & $-0.470$ \\
5 & 6.666 & $-0.113$ & 6.637 & $-0.121$ & 6.964 & $-0.347$ & 5.991 & $-0.735$ & $-0.146$ & $-0.385$ \\
\cline{2-5} \cline{6-9} \cline{10-11} 
100 & \multicolumn{4}{c}{$-154.244157$\footnotemark[6]} & \multicolumn{4}{c}{$-154.232002$\footnotemark[6]}
    & \multicolumn{2}{c}{7.627\footnotemark[7]}
\end{tabular}
\end{ruledtabular}
\footnotetext[1]{The \ccp{} and \ccpqT{} energies of the R and TS species are reported
as errors relative to CCSDT in millihartree.}
\footnotetext[2]{The \ccpqT{} values of the barrier height are reported as errors
relative to CCSDT in kcal/mol.}
\footnotetext[3]{Equivalent to CCSD.}
\footnotetext[4]{Equivalent to CCSD(T).}
\footnotetext[5]{For \%T = 1, the \ccp{} and \ccpqT{} energies obtained in
the relaxed and unrelaxed calculations are identical.}
\footnotetext[6]{Total CCSDT energy in hartree.}
\footnotetext[7]{The CCSDT barrier height in kcal/mol.}
\end{table*}

\newpage


\squeezetable
\begin{table*}[!htbp]
\caption{\label{tab:table5}
Computational timings characterizing the various CC calculations for the
cyclobutadiene/cc-pVDZ system in the transition-state (TS) geometry optimized with MR-AQCC in
Ref.\ \onlinecite{MR-AQCC}, including CCSD, CCSD(T), CR-CC(2,3), and CCSDT and the unrelaxed variants of the adaptive
\ccpq{} and \ccpqT{} approaches, abbreviated as \ccpq{}$[\%{\rm T} = x]$ and \ccpqT{}$[\%{\rm T} = x]$, respectively,
where $x = 1, 3,$ and 5,
which used the leading $x$ percent of triply excited determinants, identified on the basis
of the largest $\delta_{ijk,abc}(0)$ contributions to the CR-CC(2,3) and CCSD(T) triples corrections,
in addition to all singles and doubles, in constructing the respective $P$ spaces. The
$Q$ spaces used to determine the \ccpq{}$[\%{\rm T} = x]$ and \ccpqT{}$[\%{\rm T} = x]$ corrections
to the \ccp{}$[\%{\rm T} = x]$ energies consisted of the remaining $(100 - x)$ percent of triples not included
in the corresponding $P$ spaces.
In all post-RHF calculations, the four lowest core orbitals were kept frozen.
}
\begin{tabular}{
l @{\extracolsep{0.2in}} c @{\extracolsep{0.2in}} c @{\extracolsep{0.2in}} c @{\extracolsep{0.2in}} c}
\hline\hline
\multirow{2}{*}{Method} & \multicolumn{4}{c}{CPU time\footnotemark[1]} \\
\cline{2-5}
& $P$ Space Determination\footnotemark[2] & Iterative Steps\footnotemark[3] & Noniterative Steps\footnotemark[4]
& Total \\
\hline
CCSD & -- & 0.5 & -- & 0.5 \\
CCSD(T) & -- & 0.5 & 0.1 & 0.6 \\
CR-CC(2,3) & -- & 0.8 & 0.3 & 1.1 \\
CC($P$;$Q$)[\%T = 1] & 1.9\footnotemark[5] (2.4)\footnotemark[6] & 3.4 & 0.3
& 5.6\footnotemark[7] (6.1)\footnotemark[8]  \\
CC($P$;$Q$)[\%T = 3] & 1.9\footnotemark[5] (9.3)\footnotemark[6] & 4.0 & 0.3
& 6.2\footnotemark[7] (13.6)\footnotemark[8]  \\
CC($P$;$Q$)[\%T = 5] & 2.0\footnotemark[5] (21.3)\footnotemark[6] & 4.5 & 0.3
& 6.8\footnotemark[7] (26.1)\footnotemark[8]  \\
CC($P$;$Q$)$_\mathrm{(T)}$[\%T = 1] & 1.5\footnotemark[5] (2.0)\footnotemark[6] & 3.1 & 0.2
& 4.8\footnotemark[7] (5.3)\footnotemark[8] \\
CC($P$;$Q$)$_\mathrm{(T)}$[\%T = 3] & 1.5\footnotemark[5] (8.9)\footnotemark[6] & 3.7 & 0.2
& 5.4\footnotemark[7] (12.8)\footnotemark[8] \\
CC($P$;$Q$)$_\mathrm{(T)}$[\%T = 5] & 1.6\footnotemark[5] (20.9)\footnotemark[6] & 4.1 & 0.2
& 5.9\footnotemark[7] (25.2)\footnotemark[8] \\
CCSDT & -- & 455.8 & -- & 455.8 \\
\hline\hline
\end{tabular}
\footnotetext[1]{
All reported timings, in CPU minutes, correspond to single-core runs on the Precision 7920 workstation
from Dell equipped with Intel Xeon Silver 4114 2.2 GHz processor boards. No advantage of the
$D_{4h}$ symmetry of the TS structure of cyclobutadiene or its $D_{2h}$ Abelian
subgroup was taken in the post-RHF steps. The computational times associated with the execution of the
integral, RHF, and integral transformation and sorting routines preceding the CC steps are ignored.
}
\footnotetext[2]{
The timings associated with determining the $P$ spaces for the
unrelaxed \ccpq{}$[\%{\rm T} = x]$ and \ccpqT{}$[\%{\rm T} = x]$
computations, where $x = 1, 3,$ and 5,
include the times required to execute the corresponding initial CR-CC(2,3) and CCSD(T) runs
plus the times spent on analyzing the $\delta_{ijk,abc}(0)$ contributions to the resulting triples
corrections to CCSD needed to identify the top $x$ percent of triply excited determinants. The timings associated
with constructing the $P$ spaces for the iterative steps of the CCSD, CCSD(T), CR-CC(2,3), and
CCSDT calculations are not reported, since in these four cases these spaces are
\emph{a priori} known (consisting of all singly and doubly excited determinants in the case of
the CCSD, CCSD(T), and CR-CC(2,3) methods, and of all singly, doubly, and triply excited determinants in
the CCSDT case).
}
\footnotetext[3]{
In executing the iterative steps of each CC calculation, a convergence threshold of $10^{-7}$ hartree
was assumed. The timings corresponding to the iterative steps include the times required to
construct and solve the relevant CC amplitude equations and, in the case of CR-CC(2,3) and
\ccpq{}, the times needed to construct and solve the companion left eigenstate problems
involving the respective similarity-transformed Hamiltonians [using, in the case of
\ccpq{}, the two-body approximation discussed in the text].
}
\footnotetext[4]{
In the language of $Q$ spaces adopted by the \ccpq{} formalism, the computational times required to
determine the noniterative triples corrections of CCSD(T), CR-CC(2,3), \ccpq{}, and \ccpqT{}
correspond to all triples in the case of CCSD(T) and CR-CC(2,3) and the remaining $(100 - x)$ percent of triples
not included in the relevant $P$ spaces in the case of \ccpq{}$[\%{\rm T} = x]$ and \ccpqT{}$[\%{\rm T} = x]$
($x = 1, 3$, and 5).
}
\footnotetext[5]{
Timing obtained with the faster, but also more memory intensive, determinant selection algorithm that, in
the case of the unrelaxed \ccpq{} and \ccpqT{} computations,
first constructs and stores an array of all $\delta_{ijk,abc}(0)$ contributions to the CR-CC(2,3) or
CCSD(T) triples correction, calculated prior to the final \ccpq{}$[\%{\rm T} = x]$ or \ccpqT{}$[\%{\rm T} = x]$
run, in memory. The resulting array of all $\delta_{ijk,abc}(0)$ contributions is subsequently sorted
to identify the leading $x$ percent of triply excited determinants, for inclusion in the $P$ space needed to
set up and solve the \ccp{} equations of the target unrelaxed \ccpq{}$[\%{\rm T} = x]$ or \ccpqT{}$[\%{\rm T} = x]$
calculation, which are characterized by the largest $|\delta_{ijk,abc}(0)|$ values.
} 
\footnotetext[6]{
Timing obtained with the low-memory determinant selection algorithm that captures the leading $x$ percent
of triply excited determinants, for inclusion in the $P$ space needed to carry out the desired unrelaxed
\ccpq{}$[\%{\rm T} = x]$ or \ccpqT{}$[\%{\rm T} = x]$ calculation, on the fly. In this algorithm, when
applied to the unrelaxed computations, an array that has a dimension equal to the number of triples used
in the iterative \ccp{} steps of the target \ccpq{}$[\%{\rm T} = x]$ or \ccpqT{}$[\%{\rm T} = x]$
run is populated and continually repopulated with the
small subsets of $\delta_{ijk,abc}(0)$ contributions to the CR-CC(2,3) or CCSD(T) triples correction
as they are being successively computed, analyzed, and filtered out to identify those with the
largest $|\delta_{ijk,abc}(0)|$ values. In this procedure, the $\delta_{ijk,abc}(0)$ contribution associated
with the candidate triply excited determinant for potential inclusion in the $P$ space is compared
with the $\delta_{ijk,abc}(0)$ contribution in the array that has the smallest absolute value. If the absolute
value of $\delta_{ijk,abc}(0)$ characterizing the candidate triply excited determinant is larger than
the smallest absolute value of $\delta_{ijk,abc}(0)$ in the array, the triply excited determinant
associated with the smallest $|\delta_{ijk,abc}(0)|$ is removed from the $P$ space and replaced by the
candidate determinant. At the same time, the $\delta_{ijk,abc}(0)$ contribution in the array characterized by the
smallest absolute value is replaced by the $\delta_{ijk,abc}(0)$ corresponding to the candidate determinant and
the process is repeated by examining the next triply excited determinant for potential inclusion
in the $P$ space, and so on.
}
\footnotetext[7]{
Total CPU time resulting from the utilization of the faster, but also more memory intensive,
$P$-space determination algorithm described in footnote (e).
}
\footnotetext[8]{
Total CPU time resulting from the utilization of the low-memory $P$-space determination algorithm
described in footnote (f).
}
\end{table*}

\end{document}